\documentclass[amssymb,
reprint, superscriptaddress,longbibliography
]{revtex4-2}

\usepackage{graphicx}
\usepackage{hyperref}
\usepackage{xcolor}
\usepackage{dcolumn}
\usepackage{bm}
\usepackage[english]{babel}
\usepackage[utf8]{inputenc}
\usepackage{subcaption}
\usepackage{amsmath}
\usepackage{float}
\usepackage[mathscr]{eucal}
\usepackage{orcidlink}
\DeclareMathOperator{\Tr}{Tr}
\usepackage{physics}
\usepackage[normalem]{ulem}
\usepackage[justification=raggedright, font=small]{caption}
\usepackage{hyperref}
\hypersetup{
	colorlinks = true,
	urlcolor   = blue,
	linkcolor  = blue,
	citecolor  = cyan
}

\begin{document}

\preprint{APS/123-QED}

\title{Multiparameter quantum estimation of circulant networks}
\author{Maria Popovic{\hypersetup{hidelinks}\orcidlink{0000-0002-3297-1878}}}
\email{maria.popovic@unimi.it}
\affiliation{Dipartimento di Fisica “Aldo Pontremoli,” Università degli Studi di Milano, via Celoria 16, 20133 Milan, Italy}
\author{Silvia Cassina}%
\email{silvia.cassina@uniroma3.it}
\affiliation{Dipartimento di Scienze, Università degli Studi Roma Tre, Via della Vasca Navale 84, 00146, Rome, Italy}%
\author{Ilaria Gianani}%
\email{ilaria.gianani@uniroma3.it}
\affiliation{Dipartimento di Scienze, Università degli Studi Roma Tre, Via della Vasca Navale 84, 00146, Rome, Italy}%
\author{Claudia Benedetti{\hypersetup{hidelinks}\orcidlink{0000-0002-8112-4907}}}
\email{claudia.benedetti@unimi.it}
\affiliation{Dipartimento di Fisica “Aldo Pontremoli,” Università degli Studi di Milano, via Celoria 16, 20133 Milan, Italy}

\begin{abstract}
Networks provide a natural description of interaction patterns, transport media, and communication architectures. Accurately characterizing their  coupling structure is a key challenge in quantum technologies. We investigate the multiparameter quantum estimation of coupling strengths in circulant networks probed by continuous-time quantum walks.
We derive the quantum Fisher information matrix, identify the optimal probe states and determine the ultimate precision allowed by quantum measurements, together with its scaling with the number of unknown couplings.
We further analyze the precision achievable through position measurements of the walker, comparing the corresponding classical Cramér-Rao bounds for optimal and localized probe states with the quantum limit.
Finally, we consider finite-resolution detection modeled by coarse-grained position measurements.
Although coarse graining generally reduces estimation precision, we find that lower-resolution measurements can, in some instances, outperform other measurements with finer spatial resolution. 
Our results provide a systematic characterization of the fundamental and practical limits of coupling estimation in quantum-walk-based network inference.
\end{abstract}

\maketitle

\section{\label{sec:level1}Introduction}
Many physical  systems of interest can be represented as networks. Examples are  photonic lattices, spin networks, and quantum communication architectures \cite{Aspuru2012,Ozawa2019,Valencia26, 
Christandl2004,pemberton11, pichler15, Nokkala_2024, cirac97,kay11,Wehner2018,azuma23}. 
The functionality of these systems is largely determined by the coupling strengths between their nodes, which govern transport, communication, and information-processing capabilities. 
Accurately characterizing such couplings is therefore essential for the control, simulation, and practical implementation of networked quantum systems.
\newline
A natural approach is to probe the network through the dynamics of a quantum system evolving on its nodes and infer the couplings from the resulting measurement statistics. 
Within this framework, the inference of coupling strengths can be formulated as a Hamiltonian estimation problem \cite{Granade_2012,wiebe14,li24,Stilck24} and analysed using the tools of multiparameter quantum estimation theory \cite{Paris2008,Szczykulska2016,albarelli19,   Liu2020,Albarelli2020,Chang26}. 
Estimating several couplings simultaneously is generally harder than estimating them individually, since the optimal probe states and measurements associated with different parameters may be incompatible \cite{Ragy2016, razavian20, albarelly22,Candeloro_2021, Belliardo21,guhne23, Candeloro_2024,He_2025}. Consequently, determining the ultimate precision bounds of network inference results in a genuinely multiparameter problem.
Multiparameter quantum estimation has been  applied to the characterization of quantum Hamiltonians and networked quantum systems, ranging from quantum sensing to Hamiltonian learning and interacting many-body systems \cite{Proctor2018,eldredge18,kura18,  XIE2019, gianani23,Montenegro2024,garcia24}. 
Despite its broad applicability, exact results remain available only for a limited number of models. 
This motivates the search for physically relevant systems that admit an analytical treatment while remaining relevant to realistic multiparameter estimation scenarios.
\newline
\newline
In this work, we focus on circulant graphs, a broad class of networks characterized by discrete translational symmetry.
Circulant topologies encompass a wide range of physically relevant structures, interpolating between nearest-neighbour tight-binding models and weighted complete graphs with all-to-all interactions \cite{gray2006,Olson2014}. 
At the same time, they provide a realistic description of distance-dependent couplings while remaining analytically tractable.
Each vertex of the graph is identified with a basis state of the system while the network topology specifies the couplings between basis states. The resulting Hamiltonian governs the dynamics of quantum excitations across the network.
The coherent propagation of a quantum state  on such networks is naturally described within the framework of continuous-time quantum walks (QWs), which have become a versatile tool for the study of quantum transport \cite{plenio08,mulken2011,mulken07, kay10, razzoli11,cassio20,cavazzoni22},   quantum algorithms \cite{childs_cleve_03,childs04, hines07,portugal2013quantum,Chakraborty20,candeloro22,Radaelli_2026} and fundamental phenomena such as  quantum-to-classical transitions and decoherence \cite{schreiber11,xue13, benedetti19,gualtieri20,Bressanini22,nokkala_2026,luppi26}. Their dynamics directly encodes the underlying network structure and therefore provides a natural probe for the reconstruction and  estimation of coupling strengths \cite{ benedetti24,campbel25, Romeo_2026}.

Here, we apply multiparameter quantum estimation theory to determine the fundamental precision limits  for inferring the couplings of  circulant networks, from measurements performed on a continuous-time quantum walk.
We derive analytical expressions for the quantum Fisher information matrix for an arbitrary interaction range and show that the weak compatibility condition is always satisfied, implying that the quantum Cramér-Rao bound is attainable throughout the class of models considered.
We further characterize the optimal probe states, analyse the scaling of the quantum bound with the number of estimated couplings, and derive analytical expressions for the Fisher information associated with experimentally accessible position measurements. 
Finally, for the specific case of first- and second-neighbour couplings, we investigate the impact of finite detector resolution through  coarse-grained measurements, identifying both symmetry-induced losses of information and regimes in which lower-resolution measurements can outperform other coarse-grained measurement schemes of higher spatial resolution.
\newline
The paper is organized as follows: In Sec. \ref{sec:methods}
we introduce continuous-time quantum walks on circulant graphs and review the multiparameter estimation framework adopted throughout the work.  
In Sec. \ref{sec:parameter_compatibility} we present  analytical results about the quantum Fisher information matrix, the scaling of the quantum Cramér-Rao bound with the number of estimated parameters, and the characterization of optimal probe states.
In Sec. \ref{sec:second_neighbour_results} we specialize the analysis to the two-parameter problem corresponding to first- and second-neighbour couplings, and study both ideal and realistic measurement strategies. 
Finally, Sec.~\ref{sec:conclusion} summarizes the main results and discusses future perspectives.

\begin{figure}[]
  \centering
\begin{subfigure}[b]{1\columnwidth}
{\includegraphics[width=1\columnwidth]{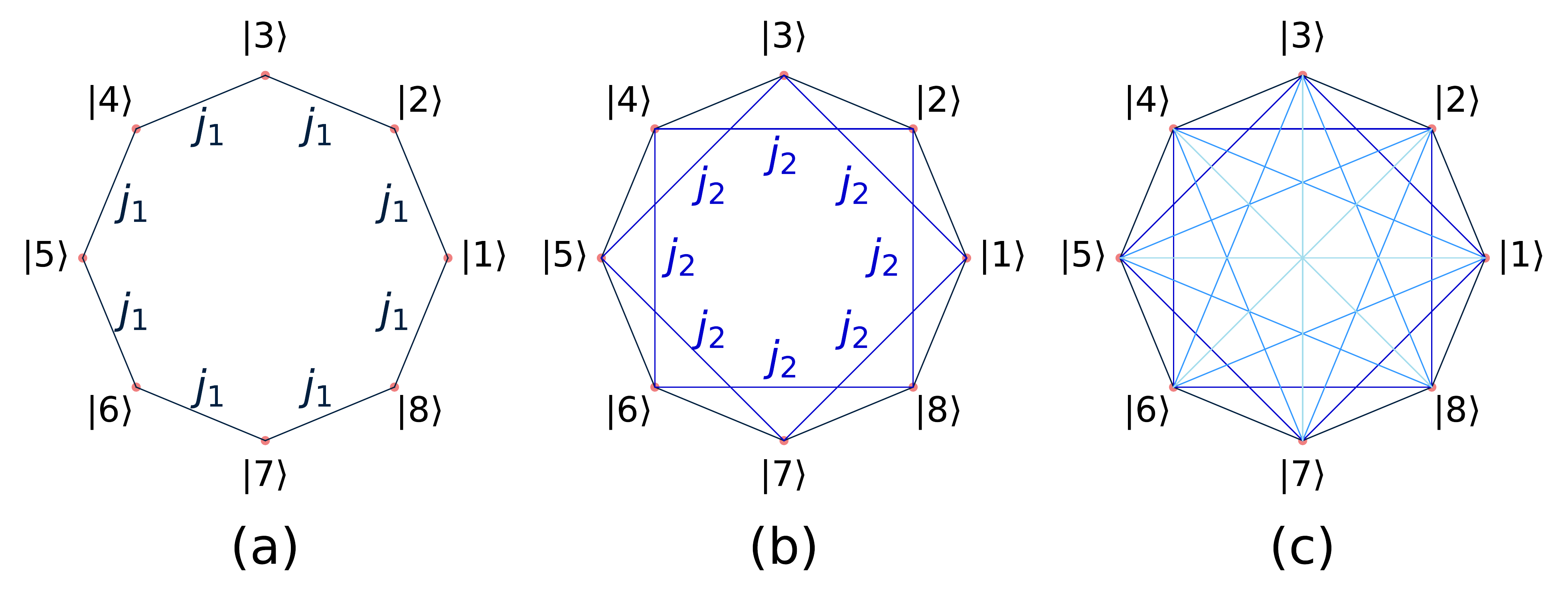}}
  \end{subfigure}
  \caption{
  Circulant graphs with $n = 8$ vertices. The interaction range between vertices can be increased while maintaining cyclic symmetry, starting from nearest‑neighbour couplings $j_1$ in (a), to second-neighbour interactions in (b), and up to a complete graph with fourth-neighbour interactions in (c). \label{fig:cyclic_graphs}}
\end{figure}

\section{Methods}\label{sec:methods}
Let $G(V,E)$ be a graph with vertex set $V$ and edge set $E$. The Hilbert space is spanned by the orthonormal position basis $\{|v\rangle\}_{v=1}^{n}$, where each basis state is associated with a graph vertex and $n=|V|$ denotes the number of vertices. We consider continuous-time quantum-walk dynamics generated by the Hamiltonian
\begin{equation}
\label{eq:qw_hamiltonian}
    \mathbf{H}  = - \sum_{(u,v) \in E}j_{uv} \left( |u \rangle \langle v | + | v \rangle \langle u |  \right) + \sum_{v \in V} \epsilon_v |v \rangle \langle v |,
\end{equation}
where $\epsilon_v$ is the on-site energy at vertex $v$, and $j_{uv} $ is the coupling strength between two connected vertices $(u, v)$. We assume symmetric couplings $j_{uv} = j_{vu}$, with $j_{uv} \in \mathbb{R}$ and  homogeneous on-site energies, $\epsilon_v = \epsilon$. 
The coupling strengths $\{j_{uv}\}$ encode the structure of the underlying network and constitute the parameters to be estimated. 
Our objective is to characterize how accurately these parameters can be inferred from measurements performed on the evolving quantum walk .
Throughout this work, we consider unitary evolutions of pure states, generated by Eq.~\eqref{eq:qw_hamiltonian}, with time-evolution operator $\mathbf{U}_t  = e^{-i t \mathbf{H}}$.
\newline
In the following, we  introduce the class of  circulant graphs, review the relevant concepts of multiparameter quantum estimation theory, and discuss how experimental constraints affect the achievable estimation precision.

\subsection{Circulant graphs} \label{subsec:circulant_graphs}
A circulant graph is a regular graph whose adjacency structure is invariant under a cyclic permutation of its vertices.
We consider circulant graphs where each vertex is connected up to its $K$-th neighbours. Increasing $K$ modifies the topology, from the simple cycle graph with only nearest-neighbour interactions, to the complete graph representing an all-to-all network, as shown in Fig.~\ref{fig:cyclic_graphs}. The coupling strengths associated with interactions at distance $\alpha$ are denoted $j_\alpha$ ($\alpha = 1,\dots,K$).

Any circulant matrix can be expressed as $\mathbf{A} = \sum_{\alpha = 1}^{n} c_{\alpha-1} \boldsymbol{\sigma}^{\alpha - 1}_n$, where $\{c_\alpha\}$ are the elements of its first row. $\boldsymbol{\sigma}_n$ is the $n \times n$ forward cyclic  shift matrix, that generates the required circulant structure, defined via its elements  $(\boldsymbol{\sigma}_n)_{kj} = \delta_{k,\,j-2 \, (\mathrm{mod}\, n)+1}$, with $k,j=1,\dots,n$.  The powers  $\boldsymbol{\sigma}_n^{\alpha}$ generate 
 jump operators that cyclically translate  the walker by $\alpha$ sites in the counter‑clockwise direction and
$\boldsymbol{\sigma}_n^0=\mathbb{I}_n$ is the identity matrix.
With this definition  each row of $\mathbf{A}$ is obtained from the previous one by a cyclic right-shift~\cite{Olson2014}.
\\
The QW Hamiltonian $\mathbf{H}$ inherits the circulant structure of the underlying graph and can thus be written in the general form $\mathbf{H} = {\epsilon \mathbb{I}_{n}} - \sum_{\alpha = 1}^{n-1} j_{\alpha}\, \boldsymbol{\sigma}^{\alpha }_n$. We set the energy offset $\epsilon$  to zero without loss of generality.
Requiring $\mathbf{H} $ to be real and symmetric,  implies $j_{\alpha}=j_{n-\alpha}$ $\forall\,\alpha=1,\dots K$, 
so that the model up to $K$-th neighbours (with $K\leq \lfloor \frac{n-1}{2}\rfloor$) takes the form:
\begin{align}
    \mathbf{H}(K) = 
    - \sum_{\alpha = 1}^{ K} j_{\alpha} (\boldsymbol{\sigma}^{\alpha}_n + \boldsymbol{\sigma}^{n - \alpha}_n )
    = \sum_{\alpha=1}^K j_{\alpha} \mathbf{H}_{\alpha} ,
    \label{circH}
\end{align}
where  $\mathbf{H}_{\alpha}=-(\boldsymbol{\sigma}^{\alpha}_n + \boldsymbol{\sigma}^{n - \alpha}_n)$ denotes the contribution associated with couplings between vertices at distance $\alpha$.
The  complete graph with an odd number of vertices is recovered from \eqref{circH}  by setting $K=(n-1)/2$. For $n$ even, an additional contribution $\mathbf{H}_{\frac{n}{2}}=-\boldsymbol{\sigma}^{ \frac{n}{2} }_n$ corresponding to the unique  vertex opposite to each site, must be included. It follows that 
\begin{align}
    \mathbf{H}_{\rm{compl}} =  \mathbf{H}\left(\lfloor\frac{n-1}{2}\rfloor\right)
-I_{\rm{even}}(n)\, j_{\frac{n}{2}} \,\boldsymbol{\sigma}^{ \frac{n}{2} }_n, 
\end{align}
where $I_{\rm{even}}(n)$ is the indicator function that returns 1 if $n$ is even and 0 otherwise. 
Since the powers $\boldsymbol{\sigma}^{\alpha}_n$  are also circulant matrices, the Hamiltonian remains circulant for any interaction range.
As such, it is diagonalized by the Fourier matrix, and its eigenvalues are~\cite{Olson2014,Seveso2019}
\begin{align}
\label{eq:eigvals1}
	\epsilon_k =   - 2 \sum_{\alpha = 1}^{ K } j_{\alpha} \cos \left( \frac{2 \pi k\, \alpha}{n}  \right)\,\,\,\, ~ \text{if } K\leq \left\lfloor \frac{n-1}{2}\right\rfloor,
\end{align}
whereas for the complete graph with even $n$,
\begin{align}
\label{eq:eigvals2}	
\epsilon_k &=  - 2 \sum_{\alpha = 1}^{ \frac{n - 2}{2}} j_{\alpha} \cos\! \left( \frac{2 \pi k \,\alpha}{n} \right) -j_{\frac{n}{2}} (-1)^{{k}}  ~ \text{if } K\! =\! \frac{n}{2}.
\end{align}
The eigenvalue index spans $k = 0, 1, \dots, n - 1$.  The eigenvectors of $\mathbf{H}$ are the columns of the Fourier matrix~\cite{Olson2014},
\begin{equation}
\label{eq:eigenstates}
	| \epsilon_k \rangle = \frac{1}{\sqrt{n}} \sum_{ v = 1}^{n} e^{ \frac{2 \pi i}{n} (v - 1) k } | v \rangle,\qquad k=0,\dots n-1
\end{equation}
and they form an orthonormal basis, satisfying $\langle \epsilon_k | \epsilon_j \rangle = \delta_{kj}$. The Hamiltonian spectrum is degenerate, with $\epsilon_k = \epsilon_{n - k}$. For even $n$, the eigenstates $|\epsilon_0 \rangle$ and $| \epsilon_{n/2} \rangle$ are the only real non-degenerate eigenvectors, while the remaining eigenvectors are degenerate and form $ (n - 2 )/2 $ complex conjugate pairs of states. For odd $n$, only the ground eigenstate $|\epsilon_0 \rangle$ is non-degenerate, and the remaining eigenvectors form $ (n - 1 )/2 $ complex conjugate pairs.

\subsection{Multiparameter estimation} \label{subsec:estimation}
Consider a  system whose state depends on a vector of unknown parameters $\mathbf{j} = \left( j_1, j_2, \dots, j_K \right)^T$. Local multiparameter estimation theory 
provides fundamental bounds on the precision with which these parameters can be estimated from measurement outcomes.
The precision of an  estimator  $\hat{\mathbf{j}} = ( \hat{j}_1, \hat{j}_2, \dots, \hat{j}_{K} )$  is quantified by its covariance matrix $\textbf{Cov}( \hat{\mathbf{j}})$, 
with elements  $\textbf{Cov}( \hat{\mathbf{j}})_{\alpha, \beta} = \langle (\hat{j}_{\alpha} - j_{\alpha}) (\hat{j}_{\beta} - j_{\beta})\rangle$, 
where the average  is taken over the conditional probability distribution $p(x|{\mathbf{j}})$ associated with  measurement outcomes. 
The multiparameter classical  Cramér-Rao (CCR) bound provides the matrix inequality
\begin{equation}
\label{eq:matrix_CRB}
    \textbf{Cov}( \hat{\mathbf{j}}) \geq \frac{1}{M}\,\mathbf{F}^{-1}_{C}, 
\end{equation}
where $M$ is the number of experimental repetitions and $\mathbf{F}_{C}$ denotes the classical Fisher information (FI) matrix, 
whose elements are 
\begin{equation}
\label{eq:classical_Fc}
    \left(\mathbf{F}_{C}\right)_{\alpha, \beta} = \sum_x\, \frac{1}{p(x | \mathbf{j})} \,\frac{\partial{p(x | \mathbf{j})} }{\partial j_{\alpha}} \frac{\partial {p(x | \mathbf{j})}}{\partial j_{\beta}  }.
\end{equation}
$\mathbf{F}_{C}$ depends explicitly on the measurement strategy through the conditional probabilities $p(x | \mathbf{j})$  of measuring outcome $x$ when the vector of parameters is fixed to $\mathbf{j}$. 
The matrix inequality Eq.~\eqref{eq:matrix_CRB} can be cast into a scalar bound by
introducing a positive weight matrix $W$ such that
$\Tr (W  \textbf{Cov}( \hat{\mathbf{j}})) \geq \Tr(W \mathbf{F}^{-1}_{C})/M $,
which allows different parameters to be weighted according to their relative importance.
In this work, we assume equal weights and set  $W = \mathbb{I}$, so that the relevant figure of merit becomes
\begin{align}
    \Tr(\textbf{Cov}( \hat{\mathbf{j}})) = \sum_{\alpha = 1}^{K} \text{Var}(\hat{j}_{\alpha}).
\end{align}
Under suitable regularity conditions, the CCR bound  is asymptotically attainable by efficient estimators in the limit of a large number of measurements~\cite{Kay1993, Paris2008, He2025March}.

In the quantum multiparameter estimation setting,  the state of the system is described by a density  operator $\hat{\rho}_{\mathbf{j} }$, depending on the vector of parameters $\mathbf{j}$. 
A general measurement is described by a positive operator-valued measure (POVM), $\{\hat{M}_{x}\}$, satisfying $\hat{M}_{x} \geq 0$ and $\sum_{x  }\hat{M}_{x} = \hat{\mathbb{I}}$  for all $x$. 
The outcome probabilities are given  by the Born rule $p(x | \mathbf{j}) = \Tr(\hat{\rho}_{\mathbf{j} }\, \hat{M}_{x})$. 
For pure states $\hat{\rho}_{\mathbf{j} } = | \psi_{\mathbf{j}} \rangle \langle \psi_{\mathbf{j}} |$, 
one finds:
\begin{equation}
\label{eq:pure_classical_FI}
   \left(\mathbf{F}_{C}\right)_{\alpha, \beta} = 4\sum_{x } \frac{ \mathrm{Re} \left( \langle \psi_{\mathbf{j} } | \hat{M}_{x} | \partial_{j_{\alpha}}\psi_{\mathbf{j} }\rangle \right) \mathrm{Re} \left( \langle \psi_{\mathbf{j} } | \hat{M}_{x} | \partial_{j_{\beta}}\psi_{\mathbf{j} }\rangle  \right)}{ \langle  \psi_{\mathbf{j} } | \hat{M}_{x} |  \psi_{\mathbf{j} }  \rangle }.
\end{equation}
Optimizing over all possible POVMs leads to the quantum Fisher information (QFI) matrix $\mathbf{F}_{Q}$~\cite{Paris2008,Liu2020}, whose elements are defined as 
\begin{equation}
\label{eq:QFI}
\left(\mathbf{F}_{Q}\right)_{\alpha, \beta} = \frac{1}{2}\Tr\left( \hat{\rho}_{\mathbf{j} } \{\hat{L}_{\alpha}, \hat{L}_{\beta} \} \right),
\end{equation}
where $L_{\alpha}$ is the symmetric logarithmic derivative (SLD),  defined implicitly through:
\begin{equation}
\label{eq:SLD}
	\partial_{j_{\alpha}}  \hat{\rho}_{\mathbf{j} }   = \frac{ \hat{L}_{\alpha}\,  \hat{\rho}_{\mathbf{j} }  +  \hat{\rho}_{\mathbf{j} }\, \hat{L}_{\alpha} }{2}.
\end{equation}
While other   QFI matrices can be defined in the multiparameter setting, throughout this work, we adopt the SLD formulation of Eq.~\eqref{eq:QFI}.
For   pure states, the QFI takes the well-known form~\cite{Liu2020}
\begin{equation}
\label{eq:pure_QFI}
	(\mathbf{F}_{Q})_{\alpha, \beta} = 4 \,\mathrm{Re} \left[ \langle \partial_{j_{\alpha}} \psi_{\mathbf{j} } | \partial_{j_{\beta}} \psi_{\mathbf{j} } \rangle - \langle \partial_{j_{\alpha}} \psi_{\mathbf{j} } |  \psi_{\mathbf{j} } \rangle \langle \psi_{\mathbf{j} } | \partial_{j_{\beta}} \psi_{\mathbf{j} } \rangle   \right].
\end{equation}
Since $\mathbf F_C\le\mathbf F_Q$ for any POVM, the scalar multiparameter quantum Cramér--Rao (QCR) bound takes the form
\begin{equation}
\label{eq:quantum_CRB}
   M\, \sum_{\alpha = 1}^{K} \text{Var}\left(\hat{j}_{\alpha}\right) \geq \Tr\left(\mathbf{F}^{-1}_{C}\right)
    \geq \Tr\left(\mathbf{F}^{-1}_{Q}\right).
\end{equation}
Unlike the single-parameter case, the multiparameter QCR bound is not generally attainable. This reflects the fact that the SLDs associated with different parameters may not commute, preventing the existence of a single optimal  measurement  that simultaneously attains the quantum limit for all parameters. When such a  optimal measurement exists, the parameters are said to be compatible.
\\
For pure states, a necessary and sufficient condition for the attainability of the SLD-QCR bound is the weak compatibility condition~\cite{Matsumoto2002,Ragy2016}:
\begin{equation}
\label{eq:weak_compatibility}
\bra{\psi_{\mathbf{j}}}\left[ {L}_{\alpha}, {L}_{\beta}\right]\ket{\psi_{\mathbf{j}}}=0\,\quad\quad\forall\,\alpha,\,\beta.
\end{equation}
For unitary parameter encoding, this condition is automatically satisfied when the dynamics is generated by mutually commuting Hamiltonian generators.~\cite{Ragy2016, Matsumoto2002}. 
\\
Since the SLD-QCR bound is not generally attainable in the multiparameter setting, the ultimate precision limit is given by the Holevo bound, which constitutes the tightest achievable bound. When the weak compatibility condition holds, the Holevo bound coincides with the SLD‑QCR bound, rendering the latter attainable~\cite{Matsumoto2002,Ragy2016,Albarelli2020}.  For pure-state model, attainability can then be achieved without requiring collective measurements, in contrast to noisy models \cite{mihai14, Roccia18}.
\newline\newline
Optimal POVMs  saturating the SLD QCR bound are, in general, not guaranteed to be experimentally feasible. 
On the other hand, projective measurements in the position basis constitute a natural and realistic measurement strategy for quantum walks. To compare their performance to the ultimate quantum limit, we introduce the  efficiency ratio~\cite{Belliardo21,Albarelli2024},
\begin{equation}
\label{eq:efficiency_ratio}
	\gamma_{\mathbf{j}} = \frac{ \Tr(\mathbf{F}^{-1}_{C})}{ \Tr(\mathbf{F}^{-1}_{Q})}.
\end{equation}
By construction, $\gamma_{\mathbf{j}} \geq 1$, with $\gamma_{\mathbf{j}}  = 1$ if and only if the QCR bound  is saturated. 
\subsection{Measurement resolution} \label{subsec:coarse_graining}

Let us consider   position measurements of a QW on a graph with $n$ vertices,  described by the projective measurement $\{ \hat{M}_{x} \}_{x = 1}^n$ with $\hat{M}_{x} = | x \rangle \langle x | $. 
In realistic experimental settings, however,  the spatial resolution of the measurement may be limited, preventing the discrimination of individual vertices. In this case, only  clusters of sites can be resolved, leading to a coarse-grained position measurement.  
We partition the graph into disjoint clusters $C_g$ of equal size $g$, i.e. each   cluster comprises $g$ vertices, such that $C_g \cap C'_g = \emptyset$ $\,\forall \,C_g $,\! $C'_g$. 
The coarse-grained POVM elements correspond to projectors onto these clusters, $\hat M_{g} = \sum_{x\in C_g} |x\rangle\langle x|$.
The resulting classical FI matrix is then given by:
\begin{equation}
\label{eq:coarse_grained_FI}
(\mathbf{F}^{g}_{C})_{\alpha, \beta} =  \sum_{C_g} \frac{1}{P \left( C_g | ~ \mathbf{j} \right)} \frac{\partial P \left( C_g | ~ \mathbf{j} \right) }{\partial j_{\alpha}} \frac{\partial P \left( C_g | ~ \mathbf{j} \right)}{\partial j_{\beta}},
\end{equation}
where the sum runs over all clusters  and  $P \left( C_g |  \mathbf{j} \right) = \sum_{x \in C_g} p(x|\mathbf{j})$ is the probability of detecting the walker within the cluster $C_g$~\cite{Razzoli2021}.
\\
For a pure state, the cluster probabilities and their derivatives read
\begin{align}
\label{eq:pure_coarse_grained_FI}
P \left( C_g | ~ \mathbf{j} \right) &= \sum_{x \in C_g} | \langle x | \psi_{\mathbf{j} } \rangle|^2,\\
\partial_{j_{\alpha}} P \left( C_g | ~ \mathbf{j} \right)  &= 2 \sum_{x \in C_g} \mathrm{Re} \left[ \langle \psi_{\mathbf{j} } |x \rangle \langle x | \partial_{j_{\alpha}} \psi_{\mathbf{j} }\rangle \right], ~\forall \alpha. \label{eq:numerator_FI_coarsed}
\end{align}
We refer to matrix $\mathbf{F}^{g}_{C}$ as the coarse-grained FI matrix associated with resolution $g$. 
\\
There are several ways a graph could be coarse-grained. In the following, we focus on the ordered clustering, in which  adjacent vertices are grouped together. 
We further assume that the cluster size $g$ divides the total number of sites $n$, ensuring a complete partition of the graph without unmeasured sites.

\section{Multiparameter Estimation of Circulant Networks}\label{sec:parameter_compatibility}
\subsection{Quantum Fisher information matrix}\label{subsec_qfim}
Let us consider the    initial state in the Hamiltonian eigenbasis of Eq~\eqref{eq:eigenstates}:
\begin{equation}
\label{eq:initial_state}
    | \psi_0 \rangle = \sum_{k = 0}^{n - 1} a_k | \epsilon_k \rangle,
\end{equation}
 with $\sum_{k = 0}^{n - 1} |a_k|^2 = 1$
and the evolved state $| \psi_t \rangle =\mathbf{U}_t\ket{\psi_0}$ 
$\ket{\psi_t}=\sum_k a_k \,e^{-it \epsilon_k}\ket{\epsilon_k}.$
\newline
Since all powers of the cyclic shift operator  commute, the Hamiltonian components satisfy 
\begin{equation}
\label{eq:commutation}
    [\mathbf{H}_{\alpha}, \mathbf{H}_{\beta}] = 0,\quad \forall \alpha,\beta = 1, \dots, K.
\end{equation}
As a consequence the QFI matrix takes the form \cite{liu15,Liu2020}:
\begin{align}
   ( \mathbf{F}_Q )_{\alpha \beta}=4\,t^2\,\text{Cov}(\mathbf{H}_{\alpha},\mathbf{H}_{\beta})_{\psi_0}
   \label{qficov}
\end{align}
where $\text{Cov}(\mathbf{H}_{\alpha},\mathbf{H}_{\beta})_{\psi_0}=\langle \mathbf{H}_{\alpha}\mathbf{H}_{\beta}\rangle_{\psi_0}-\langle \mathbf{H}_{\alpha}\rangle_{\psi_0}\langle\mathbf{H}_{\beta}\rangle_{\psi_0}$ and the expectations are taken over the initial state $\ket{\psi_0}$. Explicitly, the QFI matrix has the expression:
\begin{align}
\label{eq:qfi_matrix}
&(\mathbf{F}_{Q})_{\alpha \beta} = 16\, t^2 \left[ \sum_{k = 0}^{n - 1}  |a_k|^2 \cos \left( \frac{2 \pi k}{n} \alpha \right) \cos \left(\frac{2 \pi k}{n} \beta \right) \right. \nonumber \\ 
&\left. - \sum_{k, m = 0}^{n - 1}  |a_k|^2  |a_m|^2  \cos \left(\frac{2 \pi k}{n} \alpha \right)  \cos \left(\frac{2 \pi m}{n} \beta \right)  \right]
\end{align}
if  $\alpha, \beta \leq\left\lfloor \frac{n-1}{2}\right\rfloor$. The remaining cases, with $\alpha,\beta=\frac{n}{2}$ are reported in Appendix~\ref{appendix:qfi}, together with 
details of their derivation.
We observe that the QFI matrix is independent on the values of the coupling strengths $j_{\alpha}$	
  and depends only on the graph size $n$, the set of estimated parameters, and the populations $|a_k|^{2}$
  of the initial state in the Hamiltonian eigenbasis. Consequently, the quantum Cramér–Rao bound exhibits the well-known scaling $1/t^2$.
As demonstrated in Ref.s~\cite{Matsumoto2002, pezze17,Liu2020}, the existence of commuting generators  is a sufficient condition for the weak compatibility, Eq.~\eqref{eq:weak_compatibility}, to hold. Since the probe state is pure, this guarantees the attainability of the quantum Cramér--Rao bound.
Consequently, there exists a  projective measurement whose associated classical FI matrix coincides with the quantum FI matrix \cite{pezze17}, 
$\mathbf{F}_C(\boldsymbol{j}) = \mathbf{F}_Q(\boldsymbol{j})$ .
\newline
From the  expression in Eq. \eqref{eq:qfi_matrix}, it  follows that initial localized states on a single vertex lead to a   diagonal QFI matrix.
Indeed, a generic localized initial state $\ket{v_0}$ can be expanded in the Fourier basis as
\begin{align}
|v_0\rangle
=
\frac{1}{\sqrt n}
\sum_{k=0}^{n-1}
e^{-\frac{2\pi i}{n}(v_0-1)k}
|\epsilon_k\rangle,
\label{locstate}
\end{align}
therefore the amplitude coefficients satisfy  $|a_k|^2 = 1/n$, $\forall k$. Using the orthogonality relations of the discrete cosine functions,
$
\sum_{k=0}^{n-1}
\cos\!\left(\frac{2\pi k}{n}\alpha\right)
\cos\!\left(\frac{2\pi k}{n}\beta\right)
=
\frac{n}{2}\delta_{\alpha\beta}, 
$
valid for $1\leq\alpha,\beta \leq \lfloor (n-1)/2 \rfloor$,
together with $
\sum_{k=0}^{n-1}\cos(\frac{2\pi k\alpha}{n})=0 $
one finds that all off-diagonal elements vanish. Hence, the QFI matrix is diagonal. For odd $n$, one obtains
$(\mathbf{F}_Q)_{\alpha\beta}=8t^2\delta_{\alpha,\beta}$
whereas for even $n$, an additional diagonal element is present, $(\mathbf{F}_Q)_{\frac{n}{2}\frac{n}{2}}=4t^2$.
\subsection{Optimal initial state}
We now investigate the optimal initial state maximizing the  QCR bound, from two complementary perspectives: minimizing the single-parameter QCR, assuming all other parameters known, and optimizing the multiparameter QCR bound.
In general, these optimization problems are not equivalent as their optimal  states may not coincide.
We define the optimal initial state  $| \tilde{\psi}_0 \rangle = \sum_{k = 0}^{n - 1} \tilde{a}_k | \epsilon_k \rangle$ as the state for which the amplitude vector $\tilde{\mathbf{a}} = (\tilde{a}_0, \tilde{a}_1, \dots, \tilde{a}_{n - 1})^T$ minimizes the quantum Cramér-Rao bound. 
For the estimation of the single parameter $j_{\alpha}$, the QFI is given by the diagonal elements of the QFI matrix $(\mathbf{F}_Q)_{\alpha\alpha}=4t^2\, \text{Var}(\mathbf{H}_{\alpha})_{\psi_0}$. 
 The maximum variance of a Hermitian operator is attained by an equal superposition of its states belonging to the eigenspaces associated with the largest and smallest eigenvalues  $\epsilon_{\rm{max}}^{(\alpha)}$ and $\epsilon_{\rm{min}}^{(\alpha)}$~\cite{Popoviciu1935, Giovannetti2006}.
 In the present case,
$\epsilon_k^{(\alpha)}
=
-2\cos\!\left(\frac{2\pi k\alpha}{n}\right)$.
The minimum eigenvalue is always attained at $k_{\rm min}=0$, whereas $k_{\rm max}$ denotes the index maximizing 
$\epsilon_k^{(\alpha)}$. 
Equivalently, $k_{\rm max}$ is any integer for which $k\alpha$ is closest to $n/2$ modulo $n$. 
The  location and degeneracy of such an index depend  on their greatest common divisor $\gcd(n,\alpha)$.
The maximum information about the single parameter estimation $j_{\alpha}$ is thus extracted  by preparing the quantum walk in the state
\begin{equation}
\label{eq:general_single_parameter_opt_initial_state}
    | \tilde{\psi}_0^{(\alpha)} \rangle = \frac{1}{\sqrt{2}} \left( | \epsilon_0 \rangle +  | \epsilon_{k^{(\alpha)}_{\rm{max}}} \rangle \right).
\end{equation}
The maximal eigenvalue $k^{(\alpha)}_{\rm{max}}$ is generally degenerate. Consequently, the optimal probe state is not unique, since any normalized vector in the maximal-eigenvalue eigenspace $\mathcal{E}^{\alpha}_{\rm{max}}$ may be used in Eq. \eqref{eq:general_single_parameter_opt_initial_state}.
\newline
\newline
When estimating an arbitrary number $K$ of coupling strengths $j_1, j_2, \dots, j_K$ simultaneously, the optimization of the initial state becomes more involved. 
In general, the maximizing index $k^{(\alpha)}_{\rm{max}}$ appearing in Eq.~\eqref{eq:general_single_parameter_opt_initial_state} depends on the value of $\alpha = 1, 2, \dots K$.
Consequently, a single probe state maximizing all diagonal QFI elements $(\mathbf{F}_Q)_{\alpha\alpha}$ simultaneously may not generally exist.
Moreover,  the QFI matrix may exhibit nonvanishing off-diagonal elements, indicating correlations among the estimated parameters. 
In this case, $\Tr\left( \mathbf{F}^{-1}_{Q}\right) \neq \sum_{\alpha = 1}^K 1/(\mathbf{F}_{Q})_{\alpha \alpha}$ and the multiparameter optimization problem cannot be reduced to a collection of independent single-parameter optimizations. 
A natural figure of merit is therefore the scalar QCR bound, $\Tr\left( \mathbf{F}^{-1}_{Q}\right)$, and the optimal probe state is obtained by determining the amplitude vector $\tilde{\mathbf a}$ that minimizes this quantity. We refer to this state as the QFI-optimal  state.
\newline
\newline
A remark is in order. Whenever $\gcd(n,\alpha)=1$, i.e. $n$ and $\alpha$ are coprime, the circulant graph associated with $H_\alpha$ is isomorphic to the cycle graph generated by $H_1$. Consequently, all such generators share the same spectrum ~\cite{Coxeter1963}. 
However, this graph-theoretic equivalence does not imply the existence of a common optimal probe state. 
As shown in the next section for the case of first- and second-neighbour couplings, the probe states maximizing the individual QFI elements generally do not coincide, despite the existence of graph isomorphisms between the corresponding generators.
\subsection{Scaling with the number of  parameters $K$}
\begin{figure}[t]
\centering
\begin{subfigure}[b]{1\columnwidth}
{\includegraphics[width=0.9\columnwidth]{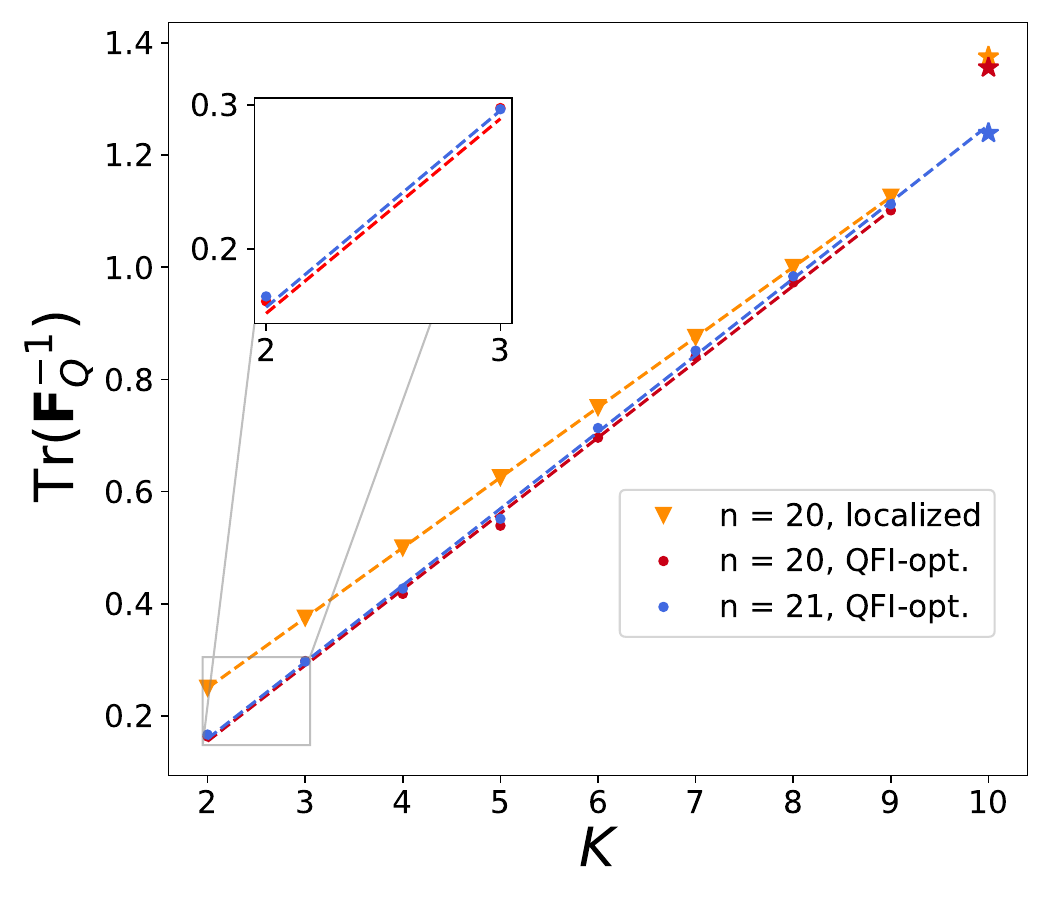}}
\end{subfigure}
\caption{ Quantum Cramér-Rao bound $\Tr(\mathbf{F}^{-1}_{Q})$ for the simultaneous  estimation of $K$ coupling strengths $j_1, \dots, j_K$. Results are shown for a localized initial state $|v\rangle=|1\rangle$ (orange triangles, $n=20$) and for numerically optimized probe states in the even ($n=20$, red circles) and odd ($n=21$, blue circles) cases. Dashed lines are linear fits. Complete-graph results are indicated by starred markers. The dimensionless evolution time is fixed at $t=1$. The inset highlights the similarity of QCR bound for different sizes $n=20$ and $n = 21$ for small values $K$.} 
\label{fig:bounds_increasing_coupling}
\end{figure}
We now investigate how the QCR bound scales with the number of estimated coupling strengths $K$. Figure~\ref{fig:bounds_increasing_coupling} compares the performances of localized initial states with QFI-optimal states, minimizing $\Tr(\mathbf{F}_Q^{-1})$, for different values of $K$.
\\
For localized initial states, the diagonal structure of the QFI matrix derived in Sec.~\ref{subsec_qfim} leads to a linear scaling of the scalar quantum Cramér--Rao bound,
\begin{equation}
\Tr(\mathbf F_Q^{-1}) = 
\begin{cases} \dfrac{K}{8\,t^2} & K\le \left\lfloor\dfrac{n-1}{2}\right\rfloor \\[2ex] \dfrac{K+1}{8t^2} \qquad\quad & K=\dfrac{n}{2} 
\end{cases} .
\label{eq:qcr_scaling_K} 
\end{equation}
The first expression includes both non-complete graphs and complete graphs with an odd number of vertices, whereas the second corresponds to complete graphs with even $n$.
Eq.\eqref{eq:qcr_scaling_K} is confirmed by the numerical results in  figure~\ref{fig:bounds_increasing_coupling}.
\\
For the  optimized initial states, no closed analytical expression for $\Tr(\mathbf{F}_Q^{-1})$ is available for a arbitrary $K$.
Moreover, numerical computations show that the corresponding QFI matrix is generally not  diagonal, indicating correlations among the estimators (with the exceptions of $K = 2$ and even $n$, see Eqs.~\eqref{eq:QFI_matrix_even_4k}-\eqref{eq:QFI_matrix_even_not4k} in Appendix~\ref{appendix:optimal_initial_state}). 
Nevertheless, Fig.~\ref{fig:bounds_increasing_coupling} shows that the optimized QCR bound retains an approximately linear dependence on $K$ for both even and odd graph sizes. 
\subsection{Position measurement}\label{sec:position_meas}
While the QCR bounds are attainable only through optimal measurements, position measurements  are directly accessible in many quantum-walk implementations and therefore provide a physically relevant benchmark.
We  consider the multiparameter  CCR bound associated with a complete position measurement, described by the POVM $\{ \hat{M}_{x} \}_{x = 1}^n$, with $\hat{M}_{x} = | x \rangle \langle x | $ and $\sum_{x = 1}^{n} \hat{M}_{x} = \hat{\mathbb{I}}_n$. 
The corresponding classical Fisher information matrix can be derived analytically for arbitrary probe states and an arbitrary number $K$ of coupling strengths, as shown in Appendix~\ref{appendix:cfi} and it can be written compactly as 
\begin{equation} 
(\mathbf{F}_C)_{\alpha\beta} = 4t^2 \sum_{x=1}^{n} \frac{ N_{\alpha,x}(\mathbf{j},t)\, N_{\beta,x}(\mathbf{j},t) }{ P_x(\mathbf{j},t) } 
\label{eq:classicalFIM}
\end{equation}
where
\begin{align*} P_x(\mathbf{j},t) &= \frac{1}{n} \sum_{k,l=0}^{n-1} \mathrm{Re} \!\left[ a_k^*a_l e^{-i\theta_{x,l,k}(\mathbf{j},t)} \right], \\ 
N_{\alpha,x}(\mathbf{j},t) &= -\frac{2}{n} \sum_{k,l=0}^{n-1} \cos\!\left(\frac{2\pi l\alpha}{n}\right) \mathrm{Im} \!\left[ a_k^*a_l e^{-i\theta_{x,l,k}(\mathbf{j},t)} \right].
\end{align*} 
Unlike the quantum FI matrix, whose elements depend only on the populations $|a_k|^2$ in the Hamiltonian eigenbasis, the classical FI matrix also depends on the coherences $a_k^*a_l$  (see Eqs.~\eqref{eq:FI_matrix_den} and \eqref{eq:FI_matrix_num}).
Moreover, the  time-dependence of the classical CR bound is determined by two distinct contributions. The first is the overall  $1/t^2$ scaling inherited from the explicit $t^2$ dependence of the classical FI matrix, analogous to the quantum Cramér-Rao bound. The second arises from the  trigonometric dependence of the FI matrix elements on the  phases
\begin{equation}
\label{eq:theta}
	\theta_{x,l,k}( \mathbf{j}, t) = t (\epsilon_l( \mathbf{j}) - \epsilon_k( \mathbf{j})) - \frac{2 \pi}{n} (x - 1) (l - k),
\end{equation}
where $x = 1, \dots, n$ labels the measured vertex, and $l, k = 0, \dots, n - 1$ are the indices of the Hamiltonian eigenvalues. 
Consequently, $\Tr(\mathbf{F}_C^{-1})$ exhibits an overall $1/t^2$ decay modulated by oscillations whose frequencies are determined by differences of Hamiltonian eigenvalues $\epsilon_l( \mathbf{j}) - \epsilon_k( \mathbf{j})$ appearing in the $\theta_{x,l,k}$ functions.
Since the elements of the classical FI matrix are constructed from trigonometric functions of these phases, the periodicity properties of  $t^2\Tr(\mathbf F_C^{-1})$  are entirely determined by the set of populated energy differences, i.e. those for which $a_l,a_k\neq 0$.
In general, these frequencies are not mutually commensurate and the resulting dynamics is  quasi-periodic. A fundamental period exists only if all populated energy differences are commensurate, i.e. if there exists a frequency $\Omega$ such that:
\begin{equation}
    \epsilon_l(\mathbf j)-\epsilon_k(\mathbf j) = m_{lk}\, \Omega, \qquad m_{lk}\in\mathbb Z,
    \label{commensurate_ene}
\end{equation} 
for all populated mode pairs $(l,k)$.
$\Omega$ represents the common fundamental frequency of the populated energy differences and may be viewed as their generalized greatest common divisor.
When Eq.\eqref{commensurate_ene} is satisfied,  all oscillatory contributions return simultaneously to their initial values after a time \begin{equation} 
\label{eq:general_period} 
T=\frac{2\pi}{\Omega}. 
\end{equation}
\section{Two-Parameter Estimation: First- and Second-Neighbour Couplings} \label{sec:second_neighbour_results}
We now focus on the specific case of a QW on a circulant graph with first- and second-neighbour couplings, depicted in Fig.~\ref{fig:cyclic_graphs}(b).
This defines a two-parameter estimation problem characterized by the parameter vector
$\mathbf j=(j_1,j_2)^T$.
Such a network requires \(n\ge 4\) vertices, with \(n=4\) corresponding to the complete graph.
We consider two different classes of initial states for the quantum walker: a   localized state and the QFI-optimal  state.
When the walker is initially localized  at vertex $v_0$, i.e.  $| \psi_0 \rangle = | v_0 \rangle$, its representation  in the Fourier basis is given by Eq.~\eqref{locstate}.
As discussed in Sec.~\ref{sec:parameter_compatibility}, all localized initial states yield the same diagonal QFI matrix, independently of the choice of $v_0$.
\subsection{Optimal initial state}\label{subsec:optimal_state}
We now consider the probe initial state that maximizes the single-parameter QFI  $(\mathbf{F}_{Q})_{\alpha\alpha}$, according to Eq.~\eqref{eq:general_single_parameter_opt_initial_state}. 
For first-neighbour couplings $j_1$, the maximizing index is \begin{align}
 k^{(1)}_{\rm{max}} =
 \begin{cases}
   {n/2} &n\,\rm{ even}\\
      {\frac{n \pm 1}{2} } &n\,\rm{ odd}
    \end{cases}
\end{align}  
where  the odd-$n$ case reflects the twofold degeneracy of the maximal eigenvalue. Consequently, any state belonging to  $\mathcal{E}_{\rm{max}}^{(1)}=\rm{span}\{|\epsilon_{\frac{n-1}{2}}\rangle,|\epsilon_{\frac{n+1}{2}}\rangle\}$ yields the same optimal value of $(\mathbf{F}_Q)_{11}$.

For second-neighbour couplings $j_2$, the maximizing indices follow from the general condition that $2k$ be closest to $n/2$ modulo $n$.
Since $ \epsilon_k^{(2)} = \epsilon_{n-k}^{(2)}$, the maximal eigenvalue of $H_2$ is  degenerate for $n>4
$. 
The maximizing indices  are 
\begin{equation} 
k_{\max}^{(2)} = \begin{cases} \left\{\dfrac{n}{4},\dfrac{3n}{4}\right\}, & n=4m, \\[2ex] \left\{\dfrac{n-1}{4},\dfrac{3n+1}{4}\right\}, & n=4m+1, \\[2ex] \left\{\dfrac{n-2}{4},\dfrac{n+2}{4}, \dfrac{3n-2}{4},\dfrac{3n+2}{4}\right\}, & n=4m+2, \\[2ex] \left\{\dfrac{n+1}{4},\dfrac{3n-1}{4}\right\}, & n=4m+3, 
\end{cases} , 
\label{eq:kmax_alpha2} 
\end{equation}
with $ m\in\mathbb{N}.$
Also in this case, any normalized state belonging to  $\mathcal{E}_{\rm{max}}^{(2)}$  yields the same optimal value of $(\mathbf{F}_Q)_{22}$.
\newline
The optimal probe states associated with the individual estimation of $j_1$ and $j_2$ do not coincide, i.e  a probe state maximizing $(F_Q)_{11}$ is not optimal for maximizing $(F_Q)_{22}$, and vice versa. While either of these states may still be used for the simultaneous estimation of both parameters, they are not expected to minimize the multiparameter QCR bound. The optimization of the individual QFI elements therefore does not solve the corresponding multiparameter estimation problem. 
\newline
\newline
We now focus on the minimization of the scalar  QCR bound, Eq.~\eqref{eq:quantum_CRB}, for the simultaneous estimation of $j_1$ and $j_2$.
The optimization of $\mathrm{Tr}(\mathbf F_Q^{-1})$ with respect to the amplitudes of the initial state was performed numerically using a sequential least-squares programming (SLSQP) algorithm. The resulting optimal probe states depend on the value of $n$ modulo four and are reported in Appendix~\ref{appendix:optimal_initial_state}. As discussed in the Appendix, the optimal probe states are supported only on the Hamiltonian eigenstates belonging to the extremal eigenspaces of $H_1$ and $H_2$.
The optimal probe state can thus be written as a linear combination of the ground state $| \epsilon_0 \rangle$, and the single-parameter  optimal initial states $| \tilde{\psi}^{(1)}_{0} \rangle$ and $| \tilde{\psi}^{(2)}_{0} \rangle$. Specifically, for even $n$,
\begin{align}
\label{eq:even_optimal_state}
    | \tilde{\psi}_0 \rangle &= e^{i \pi} \sqrt{1 - 2\,b^{2}_i(n)} |\epsilon_0\rangle \\
    & + \sqrt{2}\left(b_i(n) | \tilde{\psi}^{(1)}_{0} \rangle +  \sqrt{1 - 2\,b^{2}_i(n)} | \tilde{\psi}^{(2)}_{0} \rangle \right), \nonumber
\end{align}
where $i = 1,2$ correspond to the cases $n = 4m$ and $n \neq 4m$ ($m\in \mathbb{N}$) respectively. 
For the case $n\neq4m$, we adopt the representative single-parameter optimal probe
$| \tilde{\psi}^{(2)}_{0} \rangle = 1/\sqrt{2}\,| \epsilon_0\rangle + 1/2\,(| \epsilon_{\frac{n - 2}{4}} \rangle + | \epsilon_{\frac{n + 2}{4}} \rangle)$  which is equivalent, from the perspective of the QFI, to any state belonging to the corresponding maximal-eigenvalue eigenspace.
For odd $n$, the optimal probe state can be written as
\begin{align}
\label{eq:odd_optimal_state}
    | \tilde{\psi}_0 \rangle &= \left( b_i(n) -  c_i(n) - \sqrt{1 - b^2_i(n) -  c^2_i(n)} \right) |\epsilon_0\rangle  \\
    & + \sqrt{2} \left( \sqrt{1 - b^2_i(n) -  c^2_i(n)} | \tilde{\psi}^{(1)}_{0} \rangle + c_i(n) | \tilde{\psi}^{(2)}_{0} \rangle  \right) \nonumber,
\end{align}
where $i = 3, 4$ correspond to the cases $n -1 = 4m$ and $n -1 \neq 4m$ respectively. The functions $b_i(n)$ and $c_i(n)$ are shown in Fig.~\ref{fig:initial_amplitudes} in the Appendix. Due to the degeneracy of the relevant eigenspaces and the arbitrariness of the relative phases, the optimal initial state is not unique. Instead, the minimization of $\mathrm{Tr}(\mathbf F_Q^{-1})$ yields a family of optimal states, all leading to the same optimal QFI matrix.

Unlike the single-parameter case, the amplitudes of  the optimal multiparameter  probe  generaly depend on the graph size.  The only exception is the family of graphs with $n = 4m $  for which the optimal coefficients are independent on $n$, namely  $\tilde{a}_0 = \tilde{a}_{n/2} = b_1$ and $\tilde{a}_{n/4} = \sqrt{1 - 2\, b_1^2}$ with $b_1 \approx 0.588$ (see Eq.~\eqref{eq:optimal_initial_state_1} in Appendix \ref{appendix:optimal_initial_state}).
However, $\lim_{n \rightarrow \infty} b_i(n) = b_1$ for all $i = 2,3,4$, as shown in Fig.~\ref{fig:initial_amplitudes}. Furthermore, $\lim_{n \rightarrow \infty} c_i(n) = \sqrt{1 - 2(b_1)^2}$ for $i = 3,4$.
Substituting these asymptotic values into  Eqs.~\eqref{eq:even_optimal_state}--\eqref{eq:odd_optimal_state}, yields the large-$n$ optimal probe initial state 
\begin{align}
\label{eq:limit_optimal_state}
    | \tilde{\psi}_0 \rangle_{\!\tiny{\infty}} &= e^{i \pi} \sqrt{1 - 2\,b^{2}_1} |\epsilon_0\rangle \\
    & + \sqrt{2}\left(b_1 | \tilde{\psi}^{(1)}_{0} \rangle +  \sqrt{1 - 2\,b^{2}_1} | \tilde{\psi}^{(2)}_{0} \rangle \right). \nonumber
\end{align}
Therefore, in the large-graph limit, the amplitudes of the optimal probe become independent of the graph size. Any residual dependence on $n$ arises only through the single-parameter optimal states $|\tilde{\psi}^{(1)}_0\rangle$ and $|\tilde{\psi}^{(2)}_0\rangle$, whose structure depends on the arithmetic properties of $n$.
\newline
The corresponding optimal QFI matrices can be obtained analytically and are  reported in Appendix~\ref{appendix:optimal_initial_state}.

\subsection{Quantum Cramér-Rao bound}\label{subsec:QCR}
In this section, we compare the quantum Cramér--Rao bound obtained for a localized initial state with that achieved by  the QFI-optimal initial state, thereby assessing the impact of the probe preparation on the estimation precision.
For a localized initial state, Eq.~\eqref{locstate}, the QFI matrix is diagonal, independently of the choice of the initial vertex $v_0$, as discussed in Sec.~\ref{sec:parameter_compatibility}. The corresponding scalar quantum Cramér--Rao bound is $\Tr(\mathbf{F}^{-1}_{Q}) = \frac{1}{4\,t^2}$ for $n>4$, while $\Tr(\mathbf{F}^{-1}_{Q}) = \frac{3}{8\,t^2}$ for a complete graph ($n=4$), in agreement with Eq.~\eqref{eq:qcr_scaling_K}.

On the other hand, for the optimal probe given  in Eqs.~\eqref{eq:even_optimal_state} -\eqref{eq:odd_optimal_state}, the QFI matrix depends on the residue class of $n$ modulo four. Explicit analytical expressions for all four cases are reported in Appendix~\ref{appendix:optimal_initial_state}, Eqs.~\eqref{eq:QFI_matrix_even_4k}-\eqref{eq:QFI_matrix_odd}.
While $\mathbf{F}_Q$ is diagonal only for even $n$, all four instances converge to a common asymptotic form.
Using the large-$n$ limits of the optimal states, Eq.~\eqref{eq:limit_optimal_state}, 
one obtains:
\begin{equation}
\label{eq:qfi_matrix_asy}
   \lim_{n \rightarrow \infty}(\mathbf{F}_{Q})=  \mathbf{F}_{Q,\infty} =t^2 \begin{pmatrix}
32\, b_1^2& 0 \\
0 & 128\,(1 - 2 b_1^2)\, b_1^2 
\end{pmatrix}
\end{equation}
where $b_1\simeq0.588$.
$\mathbf{F}_{Q,\infty}$ is diagonal, implying that the estimators associated with $j_1$ and $j_2$ become asymptotically uncorrelated, and it depends only on the coefficient $b_1$ characterizing the asymptotic optimal probe state. Moreover, the convergence of the optimal amplitudes implies a universal large-$n$ metrological regime, independent of the specific arithmetic properties of the graph size.
\newline
The corresponding QCR bound  is:
\begin{equation}
\label{eq:bound_limit}
    \Tr(\mathbf{F}^{-1}_{Q,\infty}) = \frac{5 - 8\, b_1^2}{128\, b_1^2 (1 - 2\, b_1^2)\, t^2} \approx \frac{0.1636}{t^2},
\end{equation}
\begin{figure}[t]
\centering
{\includegraphics[width=0.9\columnwidth]{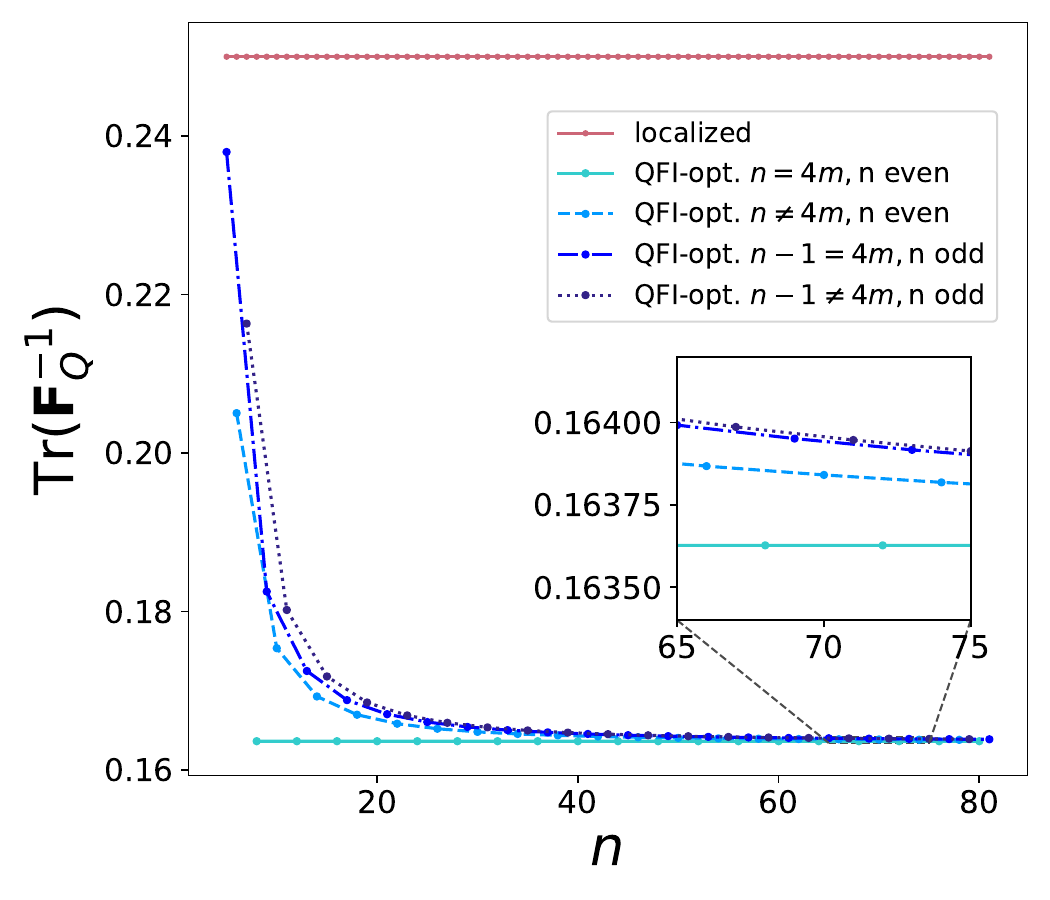}}
\caption{\raggedright Two-parameter scalar QCR bound $\Tr(\mathbf{F}^{-1}_{Q})$ as a function of the number of vertices $n$, for a localized initial state (red curve) and for the four families of the optimal initial states in Eqs.~\eqref{eq:even_optimal_state} and \eqref{eq:odd_optimal_state} (blue curves). The time is set to $t = 1$. For large $n$, all QFI-optimal-state bounds converge to the $n=4m$, $m\in\mathbb{N}$, case. The inset highlights this asymptotic convergence.} \label{fig:QFI}
\end{figure}
Fig.~\ref{fig:QFI} shows the convergence of the four classes of graphs to the asymptotic value given by  Eq.~\eqref{eq:bound_limit}, as well the behavior of the QCR bound for the  localized-state.
An interesting consequence of Eq.~\eqref{eq:qfi_matrix_asy} is that the asymptotic QFI matrix coincides with that obtained for graphs with $n=4m$ vertices. It follows that increasing the graph size beyond this regime does not further improve the ultimate estimation precision, provided that the walker is initialized in the optimal probe state.
\newline
\newline
For any graph dimension $n$, we find 
\begin{align} 
\label{eq:bound_inequality}
\Tr(\mathbf{F}^{-1}_{Q})\Big|_{\scriptscriptstyle|\tilde{\psi}_0\rangle}\, >\, \frac{1}{(\mathbf{F}_Q)_{11}\big|_{\scriptscriptstyle|\tilde{\psi}^{(1)}_0\rangle}} + \frac{1}{(\mathbf{F}_Q)_{22}\big|_{\scriptscriptstyle|\tilde{\psi}^{(2)}_0\rangle}}. 
\end{align} 
The inequality follows from the observation that for any invertible symmetric $2 \times 2$ matrix $\mathbf{A}$, $\Tr(\mathbf{A}^{-1}) = (A_{11} + A_{22})/(A_{11}A_{22} - A^2_{12} )\ge  1/A_{11} +  1/A_{22}$.    The single-parameter optimal states $|\tilde{\psi}^{(1)}_0\rangle$ and $|\tilde{\psi}^{(2)}_0\rangle$ maximize the variances of $H_1$ and $H_2$, respectively. 
Since the multiparameter-optimal states of Eq.s \eqref{eq:even_optimal_state}--\eqref{eq:odd_optimal_state} are nontrivial superpositions of the single-parameter optimal states, they cannot maximize the diagonal multiparameter QFI entries individually.
Hence $(\mathbf{F}_Q)_{11}\big|_{\scriptscriptstyle|\tilde{\psi}_0\rangle} < (\mathbf{F}_Q)_{11}\big|_{\scriptscriptstyle|\tilde{\psi}^{(1)}_0\rangle} $ and $(\mathbf{F}_Q)_{22}\big|_{\scriptscriptstyle|\tilde{\psi}_0\rangle}  < (\mathbf{F}_Q)_{22}\big|_{\scriptscriptstyle|\tilde{\psi}^{(2)}_0\rangle} $, thus satisfying Eq.~\eqref{eq:bound_inequality}.
The right-hand side of Eq.~\eqref{eq:bound_inequality} corresponds to the  scenario in which $j_1$ and $j_2$ are estimated independently, each using its own single-parameter optimal probe state. The  inequality therefore quantifies the loss in precision resulting from the requirement that both parameters be estimated simultaneously using a single probe state.
\begin{figure}[t]
  \centering
  \begin{subfigure}[b]{0.5\linewidth}
    \includegraphics[width=\linewidth]{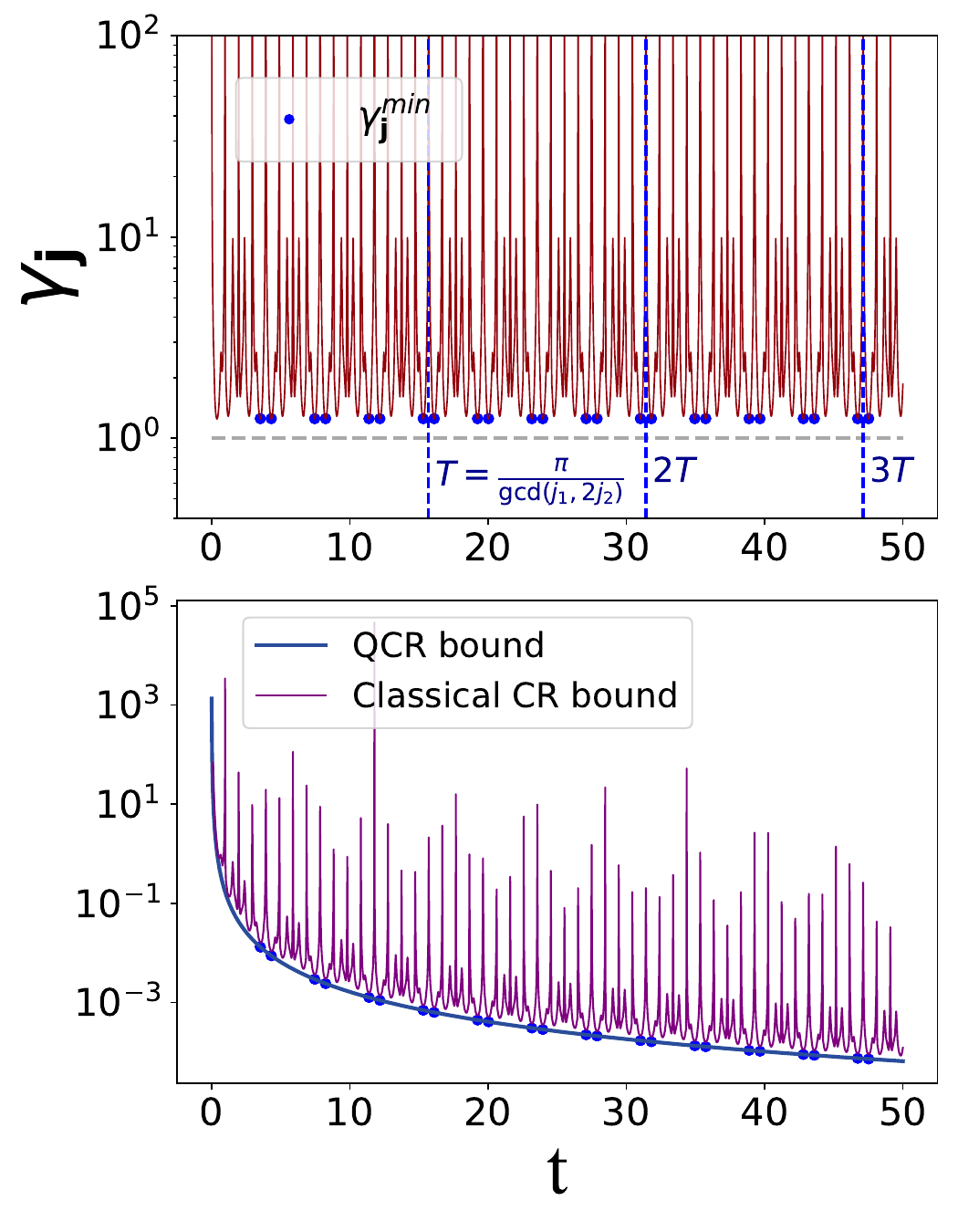}
    \caption{$n = 8$}
  \end{subfigure}\hfill
  \begin{subfigure}[b]{0.5\linewidth}
    \includegraphics[width=\linewidth]{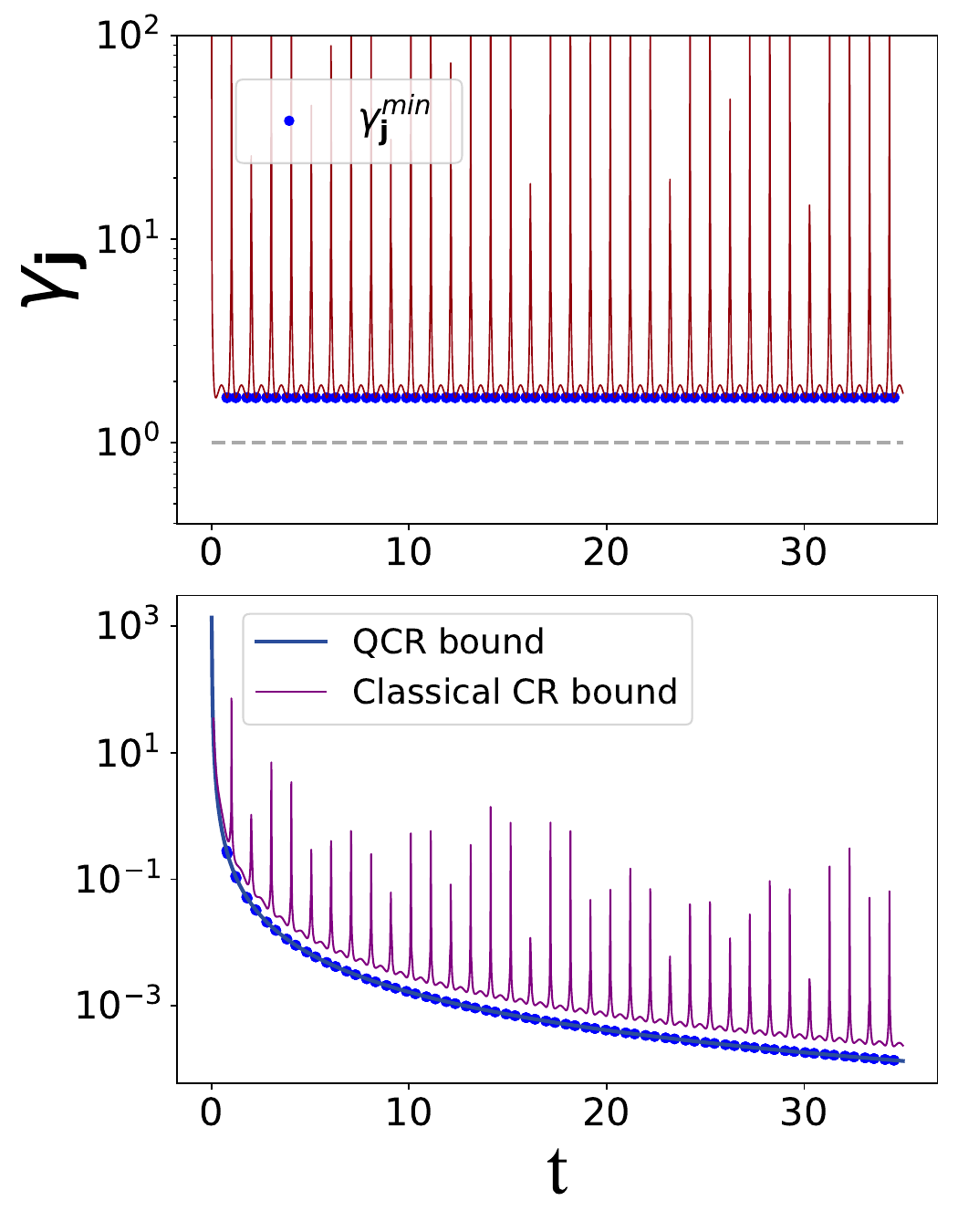}
    \caption{$n = 41$}
  \end{subfigure}
  \vspace{4pt} 
  \begin{subfigure}[b]{0.5\linewidth}
    \includegraphics[width=\linewidth]{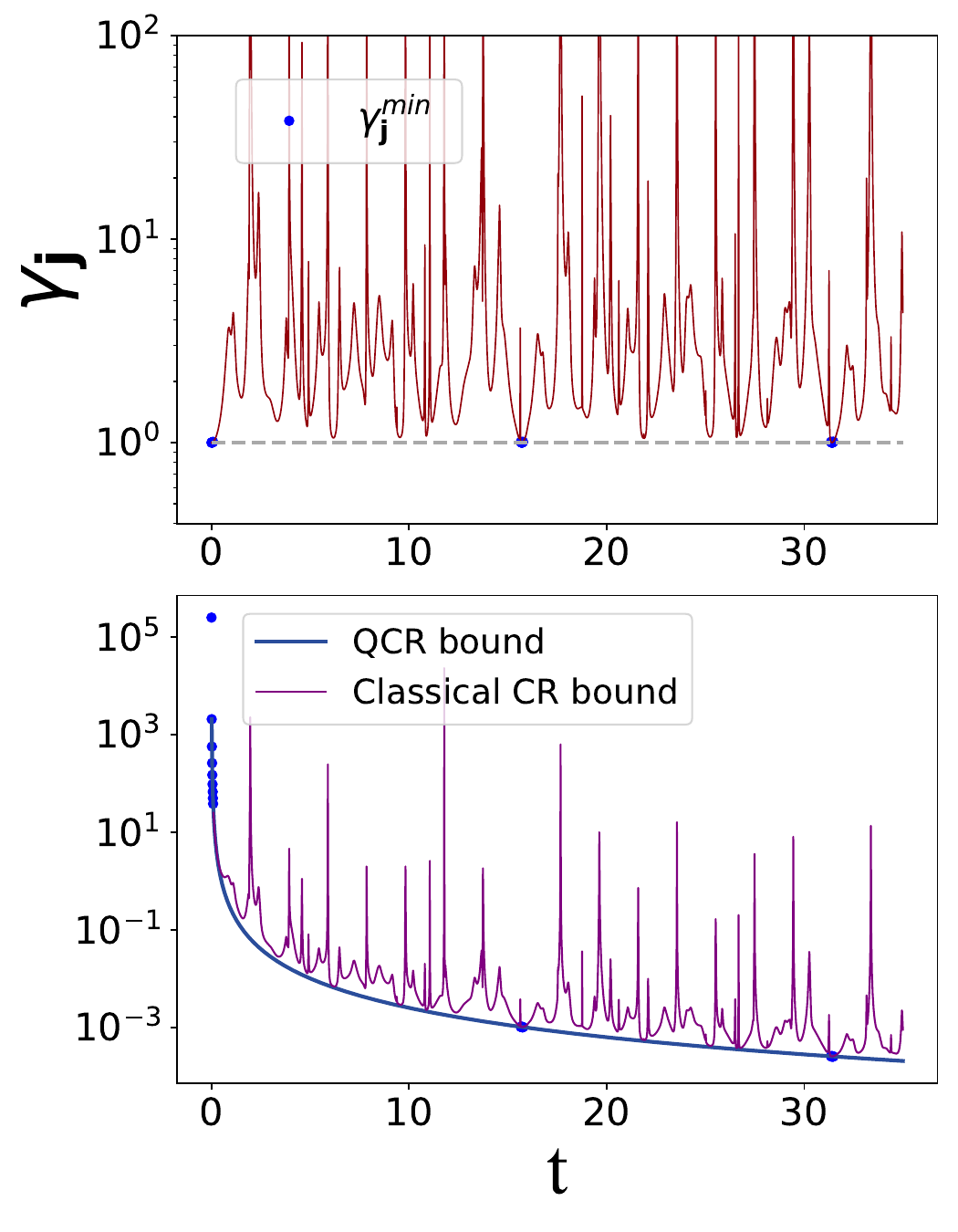}
    \caption{$n = 8$}
  \end{subfigure}\hfill
  \begin{subfigure}[b]{0.5\linewidth}
    \includegraphics[width=\linewidth]{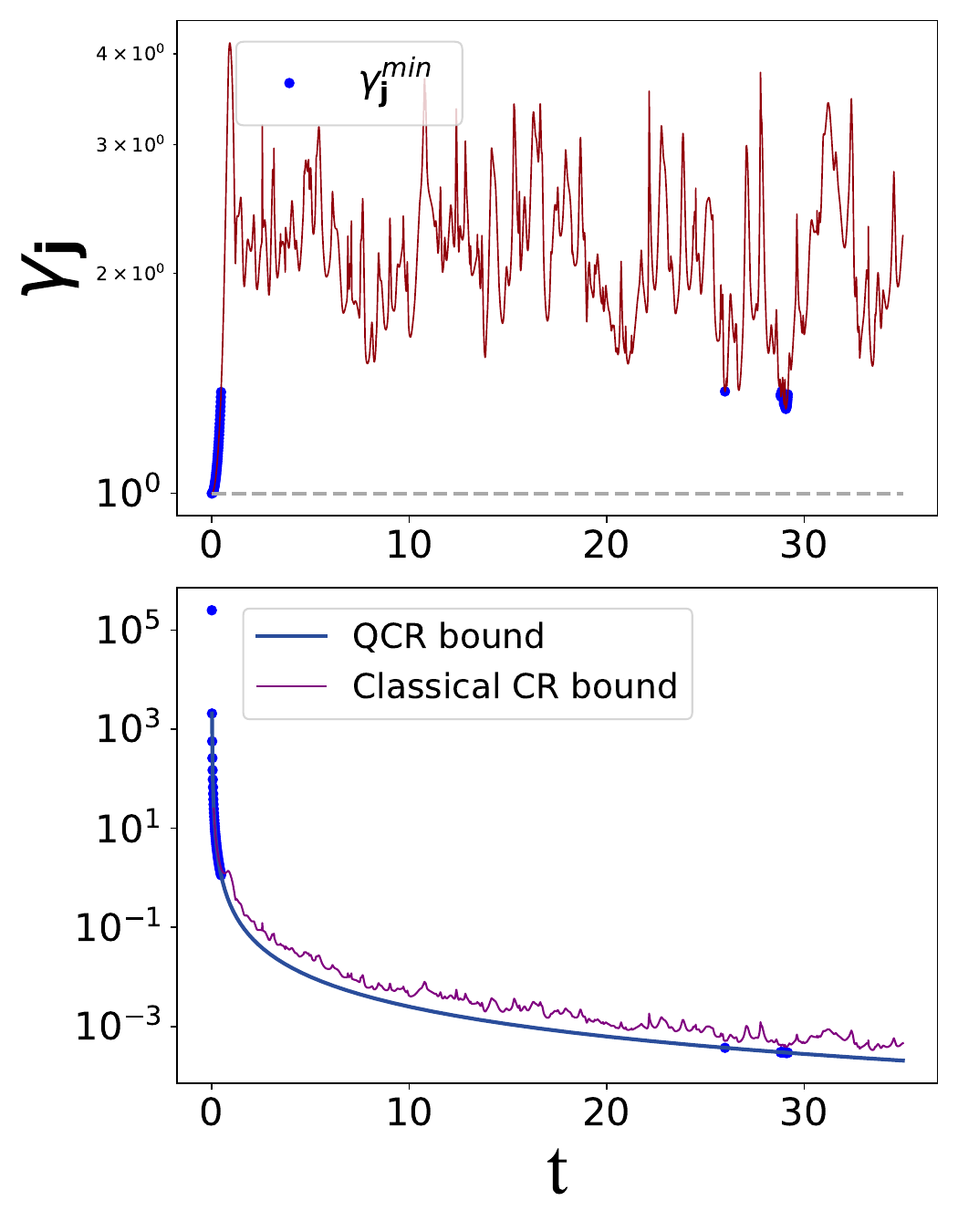}
    \caption{$n= 41$}
  \end{subfigure}
\caption{Efficiency ratio $\gamma_{\mathbf j}$ (red lines) and the corresponding classical and quantum Cramér-Rao bounds (purple and blue lines, respectively) for complete position measurements. The blue markers identify the values of $\gamma_{\mathbf j}$ lying within a distance $\Delta$ of the global minimum in the considered time interval. The horizontal gray dashed line indicates the optimal value $\gamma_{\mathbf j}=1$. Coupling strengths are set to $j_1=1$ and $j_2=0.8$. Panels (a) and (b) correspond to QFI-optimal probe states, while panels (c) and (d) correspond to an initial state localized at $|v_0\rangle=|1\rangle$. In panels (a)-(c), $\Delta=0.01$, whereas in panel (d), $\Delta=0.38$. The vertical dashed lines in panel (a) indicate the exact oscillation period $T$ for the $n=8$ case.}
  \label{fig:time_behaviour}
\end{figure}
\subsection{Complete position measurement}\label{subsec:complete_measurement}
\begin{figure}[t]
\centering
{\includegraphics[width=0.9\columnwidth]{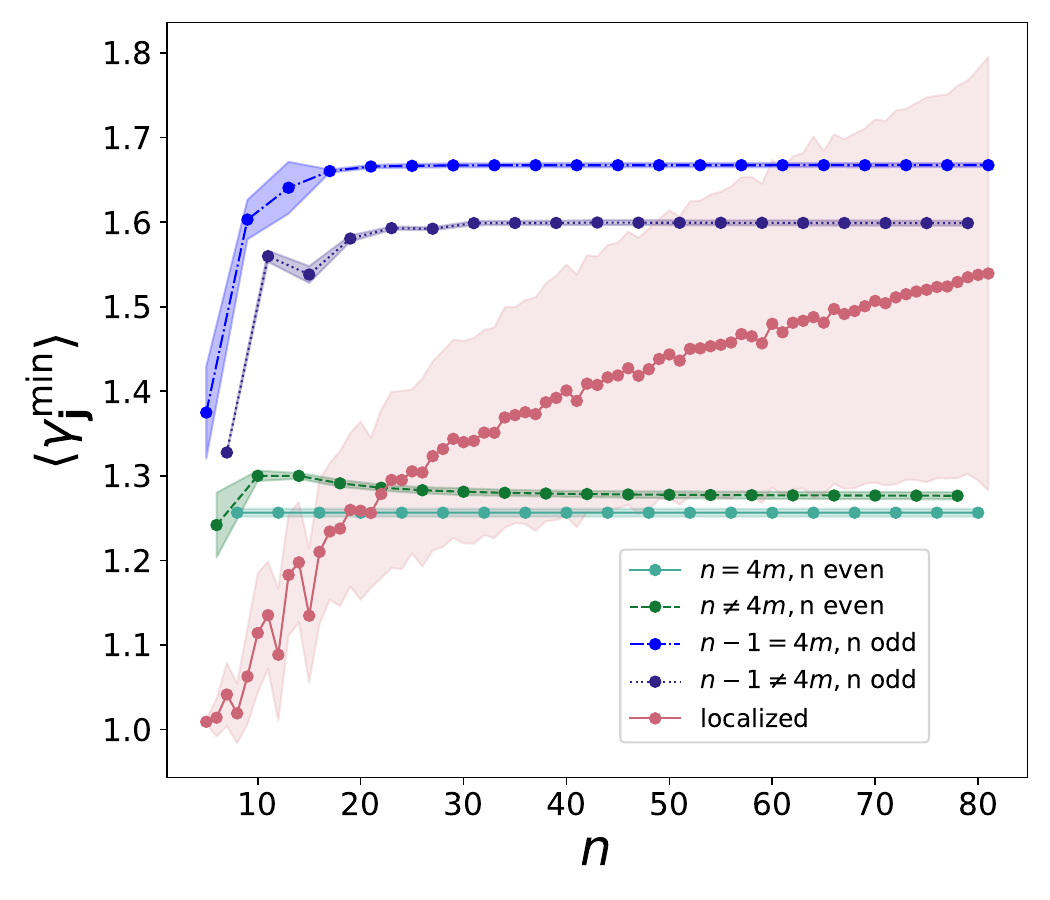}}
\caption{\raggedright 
Average optimal efficiency ratio $\langle \gamma_{\mathbf j}^{\rm min}\rangle$ for complete position measurements in the time interval $[0.001,50]$,  as a function of the number of vertices $n$. The four families of QFI-optimal probe states, Eqs.~\eqref{eq:optimal_initial_state_1}--\eqref{eq:optimal_initial_state_4}, are represented by the light-green, green, light-blue, and blue markers, respectively, while red markers correspond to the localized probe state $|v_0\rangle=|1\rangle$. For each value of $n$, markers indicate the average over $10^4$ coupling pairs $(j_1,j_2)$ sampled uniformly from $(0,1]$, and the shaded bands denote one standard deviation. The quantity $\gamma_{\mathbf j}^{\rm min}$ is computed using a time step $\delta t=0.01$ and a tolerance $\Delta=0.01$ with respect to the global minimum in the considered time interval.
}
\label{fig:CFI}
\end{figure}

To assess the performance of complete position measurements relative to the ultimate quantum limit,  we analyse the efficiency ratio $\gamma_{\mathbf{j}}$, defined in Eq.~\eqref{eq:efficiency_ratio}.
The comparison is carried out for both localized initial states and the QFI-optimal probe states given by Eqs.~\eqref{eq:even_optimal_state}--\eqref{eq:odd_optimal_state}.
Since both $\Tr(\mathbf{F}^{-1}_C)$ and  $\Tr(\mathbf{F}^{-1}_Q)$ exhibit the same asymptotic $1/t^2$ scaling, this contribution  cancels in the $\gamma_{\mathbf{j}}$. 
The temporal behaviour of $\gamma_{\mathbf j}$, therefore, is governed by the phase factors $\theta_{x,l,k}(\mathbf j,t)$ in Eq.~\eqref{eq:theta}.
Since the efficiency ratio inherits its time dependence from the classical FI matrix, its temporal behaviour is periodic or quasi-periodic according to the conditions discussed in Sec.~\ref{sec:position_meas}.
For the circulant quantum walks considered here, the commensurability condition is generally not satisfied. An important exception occurs for the QFI-optimal probe states on graphs with $n=4m$, $m\in\mathbb N$. 
As shown in Appendix~\ref{appendix:cfi}, for rational coupling strengths $j_1$ and $j_2$ the rescaled classical CR bound $t^2\Tr(\mathbf F_C^{-1})$ is exactly periodic, with period 
\begin{equation} 
\label{eq:n4_period} 
T=\frac{\pi}{\gcd(j_1,2j_2)}. 
\end{equation}
This expression corresponds to Eq.~\eqref{eq:general_period} for the set of energy modes populated by the optimal state. For all other graph sizes and initial states considered here, both $t^2\Tr(\mathbf F_C^{-1})$ and  $\gamma_{\mathbf j}$, exibit quasi-periodic dynamics.
\newline
\newline
The oscillatory time dependence of the efficiency ratio is shown in Fig.~\ref{fig:time_behaviour} for both  QFI-optimal and a localized initial state. 
Consistent with the discussion above, the $n=8$ optimal-state case shown in Fig.~\ref{fig:time_behaviour}(a) exhibits strictly periodic oscillations, whereas all other examples display quasi-periodic dynamics.
In all cases, the efficiency ratio shows a sequence of  local minima at which the optimal value $\gamma_{\mathbf j}=1$ is approached. These minima correspond to measurement times for which a complete position measurement performs closest to the optimal POVM.
However, a value $\gamma_{\mathbf j}\simeq1$ does not necessarily imply highly precise parameter estimation.
As illustrated in Fig.~\ref{fig:time_behaviour}, both the classical and quantum Cramér-Rao bounds decrease asymptotically as $1/t^2$. 
Consequently, even when the efficiency ratio is close to unity, both bounds may remain large at short times, resulting in poor estimation precision. 
In practice, one therefore seeks a measurement time that combines a small classical Cramér-Rao bound, corresponding to high estimation precision, with a near-optimal efficiency ratio.
To quantify this trade-off within a finite experimental interval $[t_0,t_f]$,  we introduce a tolerance parameter $\Delta$
and  define the set of relevant local minima, represented by the blue markers in Fig.~\ref{fig:time_behaviour}, as the set of values of $\gamma_{\mathbf{j}}$ lying within a distance $\Delta$  from the global minimum, $\min_{t \in [t_0, t_f] }\gamma_{\mathbf{j}} (t)$, attained in the time-interval.
Among these minima, we select the one occurring at the latest  time within the interval and define the corresponding optimal measurement time $\tau$ as
\begin{equation}
\label{eq:optimal_time}
    \tau = \max \left\{ t' \! \in [t_0, t_f]\!:\! \left|\gamma_{\mathbf{j}} (t') -\! \min_{t \in [t_0, t_f]}\!\gamma_{\mathbf{j}} (t)\right| \leq \Delta \right\}.
\end{equation}
The corresponding best achievable efficiency ratio within the interval $[t_0, t_f]$ is then
\begin{equation}
    \gamma^{\rm{min}}_{\mathbf{j}} = \gamma_{\mathbf{j}}(\tau),
\end{equation}
which quantifies the closest practically achievable approximation to the quantum limit within the available experimental time window. 
For localized initial states, particularly on large graphs, the first minima of $\gamma_{\mathbf j}$ may occur at very short times (see Fig.~\ref{fig:time_behaviour}(d)). This makes the optimization criterion introduced above especially relevant.
\newline
\newline
Figure~\ref{fig:CFI} shows the averaged value $\langle \gamma^{\min}_{\mathbf j}\rangle$, together with the corresponding spread over $10^4$ randomly sampled coupling pairs $(j_1,j_2)$, as a function of the number of vertices $n$ and for different initial states.
For the QFI-optimal probe states, $\gamma^{\min}_{\mathbf j}$ exhibits only a weak dependence on the coupling strengths, as indicated by the small spread observed for all graph sizes. In contrast, the localized initial state shows a much broader distribution, with variability increasing with $n$, revealing a stronger coupling dependence and a progressive degradation of the measurement efficiency. 
\newline
For the QFI-optimal probe state $|\tilde{\psi}_0\rangle$ defined in Eqs.~\eqref{eq:even_optimal_state}-\eqref{eq:odd_optimal_state}, the averaged efficiency ratio satisfies $\langle \gamma_{\mathbf j}^{\rm min}\rangle>1$ for all graph sizes.
This indicates that complete position measurements are  suboptimal compared with the optimal POVM, although the associated efficiency loss is smaller for small graphs. 
As $n$ increases, the optimal-state curves rapidly converge to four distinct asymptotic values, determined by whether $n$ (or $n-1$ for odd graphs) is divisible by four, while the localized-state behaviour remains qualitatively different.
\newline
\newline
A  remarkable feature is observed for the family $n=4m$, whose asymptotic value remains independent of the graph size. This property is analogous to that observed for the quantum Cramér-Rao bound and originates from the structure of the classical Fisher information matrix for the QFI-optimal probe states. 
For $n=4m$, the explicit expressions reported in Appendix~\ref{appendix:cfi} show that both $P_s(\mathbf j,t)$ and $N_{\alpha,s}(\mathbf j,t)$ depend on the vertex index $s$ only through trigonometric functions of $\pi s/2$ and $\pi s$. 
Consequently,  $P_{s+4}(\mathbf j,t)=P_s(\mathbf j,t)$,  $ N_{\alpha,s+4}(\mathbf j,t)=N_{\alpha,s}(\mathbf j,t),$  and thus the summand in Eq.~\eqref{eq:classicalFIM},  $g_{\alpha\beta}(s) = \frac{ N_{\alpha,s}(\mathbf j,t) N_{\beta,s}(\mathbf j,t) }{ P_s(\mathbf j,t) }$,  is periodic in $s$ with period four. 
Each element of the Fisher information matrix can therefore be written as the sum of  period-four functions, $(\mathbf F_C)_{\alpha\beta} = \frac{4t^2}{n} \sum_{s=1}^{n} g_{\alpha\beta}(s)$.
When $n=4m$, \[ \frac{1}{4m} \sum_{s=1}^{4m} g_{\alpha\beta}(s) = \frac14 \sum_{s=1}^{4} g_{\alpha\beta}(s), \] so every element of $\mathbf F_C$  is independent of the graph size. As a consequence, both the classical Cramér-Rao bound $\Tr(\mathbf F_C^{-1})$ and the corresponding efficiency ratio $\gamma_{\mathbf j}$ are independent of $n$ for QFI-optimal probe states.

\subsection{Coarse-grained position measurement}
We now investigate the precision attainable when the position of the walker is measured with finite spatial resolution.
As discussed in Sec.~\ref{subsec:coarse_graining}, a coarse-grained position measurement resolves clusters  $C_g$ composed of $g$ graph vertices, rather than individual sites. Throughout this section we assume all clusters have equal size.
The corresponding  multiparameter CCR bound $\Tr[(\mathbf{F}^{g}_{C})^{-1}]$ is determined by the coarse-grained FI matrix  defined in Eq.~\eqref{eq:coarse_grained_FI}.  
\newline
Our first objective is to identify the conditions under which coarse-grained measurements contain sufficient information to estimate the couplings $\mathbf j=(j_1,j_2)$.  
For an initially localized state  $|\psi_0 \rangle = | v_0 \rangle$, numerical evidence indicates that the matrix $\mathbf{F}^{g}_{C}$ is generically invertible once the measurement time exceeds a threshold 
that increases with the cluster size $g$.
This behaviour reflects the finite propagation of information away from the cluster containing the initially occupied vertex. At sufficiently short times, coarse-graining may therefore prevent the simultaneous estimation of the two couplings.
\newline
For the QFI-optimal initial state $| \tilde{\psi}_0 \rangle$ of  Eqs.\eqref{eq:even_optimal_state}-\eqref{eq:odd_optimal_state}, this propagation delay is absent because the probe is delocalized over the entire graph. 
However,  parameter estimation may be prevented by other factors. 
In particular, when the graph size is $n = 4m$ and the ordered clusters have size $g=4m'$, with $m,\,m' \in \mathbb{N}$ the coarse-grained measurement becomes completely insensitive to the parameters and the the coarse-grained FI matrix elements are always zero.
As shown in Appendix~\ref{appendix:coarse_grained},
the site probabilities satisfy the period-four relation $ P_{x + 4}(\mathbf{j},t)= P_x(\mathbf{j},t)$ at all times.
Consequently, each ordered cluster contains an integer number of complete spatial periods and its probability is $P(C_k|\mathbf j)=\frac{g}{n}$, independent of $j_1$, $j_2$, and $t$. The coarse-grained probability distribution therefore carries no information about the parameters, implying $\mathbf F_C^g=\mathbf 0 $.
This cancellation mechanism relies on the ordered structure of the clustering. For non-ordered clusterings, in which each cluster consists of random non-adjacent vertices, the periodic contributions do not generally cancel and $\mathbf F_C^g$ is typically invertible. 
Likewise, outside the symmetry-protected case $n=4m$ and $g=4m'$, our numerical analysis indicates that the ordered-coarse-grained FI matrix is generically invertible.

Further numerical evidence suggests that the QFI-optimal probes may become ineffective under coarse-grained detection. For $g > 1$, the minima of $\gamma_{\mathbf{j}}$ are one to two orders of magnitude larger than those for a complete measurement of  Fig.~\ref{fig:time_behaviour}, so coarse-graining generally operates far from the QFI limit. 
Consequently, optimizing the probe with respect to the QFI yields limited practical advantages.
In the remainder of this section we  focus on localized initial states, which are experimentally more accessible and remain informative under coarse graining.
\newline
For localized probes, the coarse-grained FI matrix exhibits a simple translation symmetry with respect to the initial vertex.
 Consider an ordered clustering where each cluster, of  fixed size $g$, is   $C_r=\{r, r+1,\dots,r+g-1\}$. We can write the cluster probability as
 \begin{equation}
 P(C_r|\mathbf j) = \frac{1}{n^2} \sum_{l,k=0}^{n-1} F_{l,k}(\mathbf j,t)\, e^{-i(l-k)\,\zeta_{v_0,r}}, 
 \end{equation} 
 where 
 \begin{align} 
 &\zeta_{v_0,r} = \frac{2\pi}{n} \left[v_0-1-(r-1)g\right],\\
 &F_{l, k}(\mathbf{j},t) = \sum_{z=1}^{g} e^{-i\theta_{z,l,k}(\mathbf{j},t)}.
 \end{align}
Since, $\zeta_{v_0, r + 1} = \zeta_{v_0 - g, r} $, it follows that
\begin{align}
P\left(C_{r + 1}\,\middle|\,\mathbf{j}\right)[v_0] = P(C_r|\mathbf j)[(v_0-g)\!\!\!\mod n],
\end{align}
with the cluster index taken modulo $n/g$. 
The $\mathbf{j}$-derivatives only act on $F_{l, k}(\mathbf{j},t)$, so the translation symmetry holds for  all elements of the coarse-grained FI matrix in Eq.~\eqref{eq:coarse_grained_FI}, which we rewrite compactly as $(\mathbf{F}^{g}_{C})_{\alpha, \beta}[v_0] = \sum_{r = 1}^{n/g} f_{\alpha, \beta} (r, v_0)$ for ordered clustering. Owing to the translation symmetry, $ f_{\alpha, \beta} (r, v_0 - g) = f_{\alpha, \beta} (r + 1, v_0)$, with $r + 1$ taken modulo $n/g$, and $v_0 - g$  modulo $n$. 
Summing over all clusters and relabelling the cluster index  gives:
\begin{equation}
   \mathbf F_C^g[(v_0-g)\!\!\!\mod n] = \mathbf F_C^g[v_0].
\end{equation}
The coarse-grained Cramér--Rao bound is periodic in the initial vertex with period $g$.
Optimizing the estimation protocol over the initial site therefore requires exploring only the $g$ vertices of a single cluster, rather than all $n$ vertices of the graph.
\newline
\newline
To quantify the loss of precision due to finite resolution, we consider the relative increase of the Cramér--Rao bound,
$\delta_g=\left[\Tr(\mathbf{F}_{C}^{g})^{-1}) -\Tr(\mathbf{F}^{-1}_{C})\right]/ \Tr(\mathbf{F}^{-1}_{C})$, with respect to the complete position measurement. 
\begin{figure}[t]
\centering
{\includegraphics[width=0.8\columnwidth]{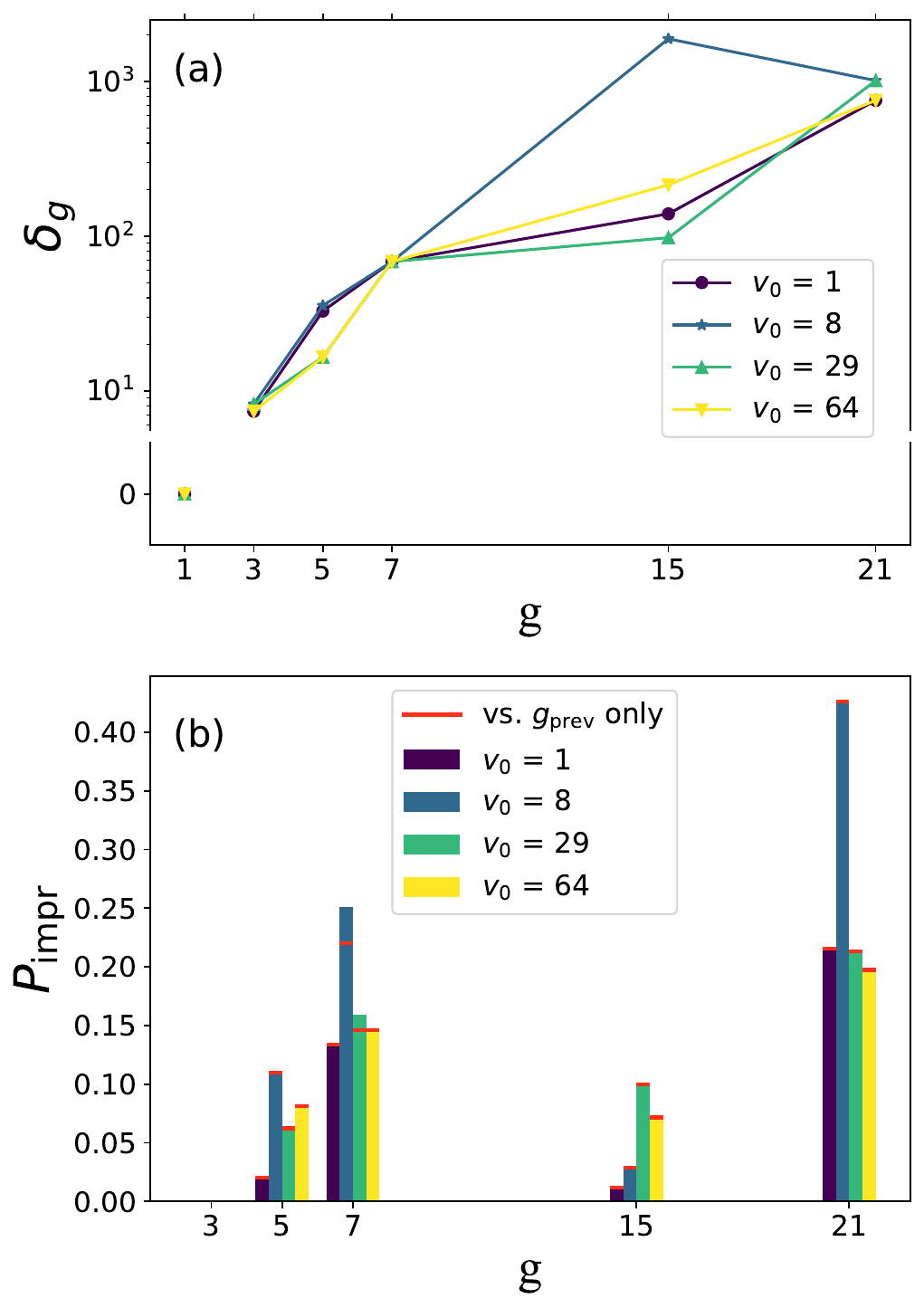}}
\caption{ (a): Relative distance between the coarse-grained CR bound and the complete measurement CR bound, as a function of the cluster size $g$, for a single graph realization with $j_1 = 1$ and $j_2 = 0.8$. Four localized initial states $|v_0\rangle$ are shown, illustrating that the dependence on $g$ can be non-monotonic.
(b):  Probability that a measurement with cluster size $g$ leads to a smaller CR bound than a measurement with larger resolution $g' < g$, for the same initial states.
The red curves show the probability of outperforming the immediately finer resolution, $ g' = g_{\rm prev}$, where $g_{\rm prev}$ is the divisor of $n$ immediately preceding $g$. Probabilities are computed over $10^4$ graph realizations with different $(j_1, j_2)$ couplings. Measurement time is fixed to $t = 20$ and the graph size  to $n = 105$.  \label{fig:coarsed}}
\end{figure}
In Fig.~\ref{fig:coarsed}(a) we show that, for a single graph realization,  coarse graining can significantly degrade the achievable precision. 
More interestingly, the degradation is not necessarily monotonic in the cluster size. 
Although every coarse-grained measurement ($g>1$) performs worse than the complete position measurement ($g=1$), there exist values of $g$ for which the corresponding bound is smaller than the one obtained with a higher spatial-resolution measurement.
To quantify the prevalence of this effect, we define the probability $P_{\rm impr}$ that a coarse-grained measurement with cluster size $g$ yields a smaller CCR bound than at least one measurement with cluster size $g'<g$. As shown in Fig.~\ref{fig:coarsed}(b), this probability can be substantial, reaching values of order $40\%$ for $n=105$.
In most cases, the improvement is observed only with respect to the closest finer resolution, corresponding to the divisor of $n$ immediately preceding $g$, see
 the red bars in Fig.~\ref{fig:coarsed} (b). Thus, while reducing the measurement resolution can  improve precision, such improvements generally remain distant from the performance of the complete position measurement.

\begin{figure}[t]
\centering
{\includegraphics[width=0.8\columnwidth]{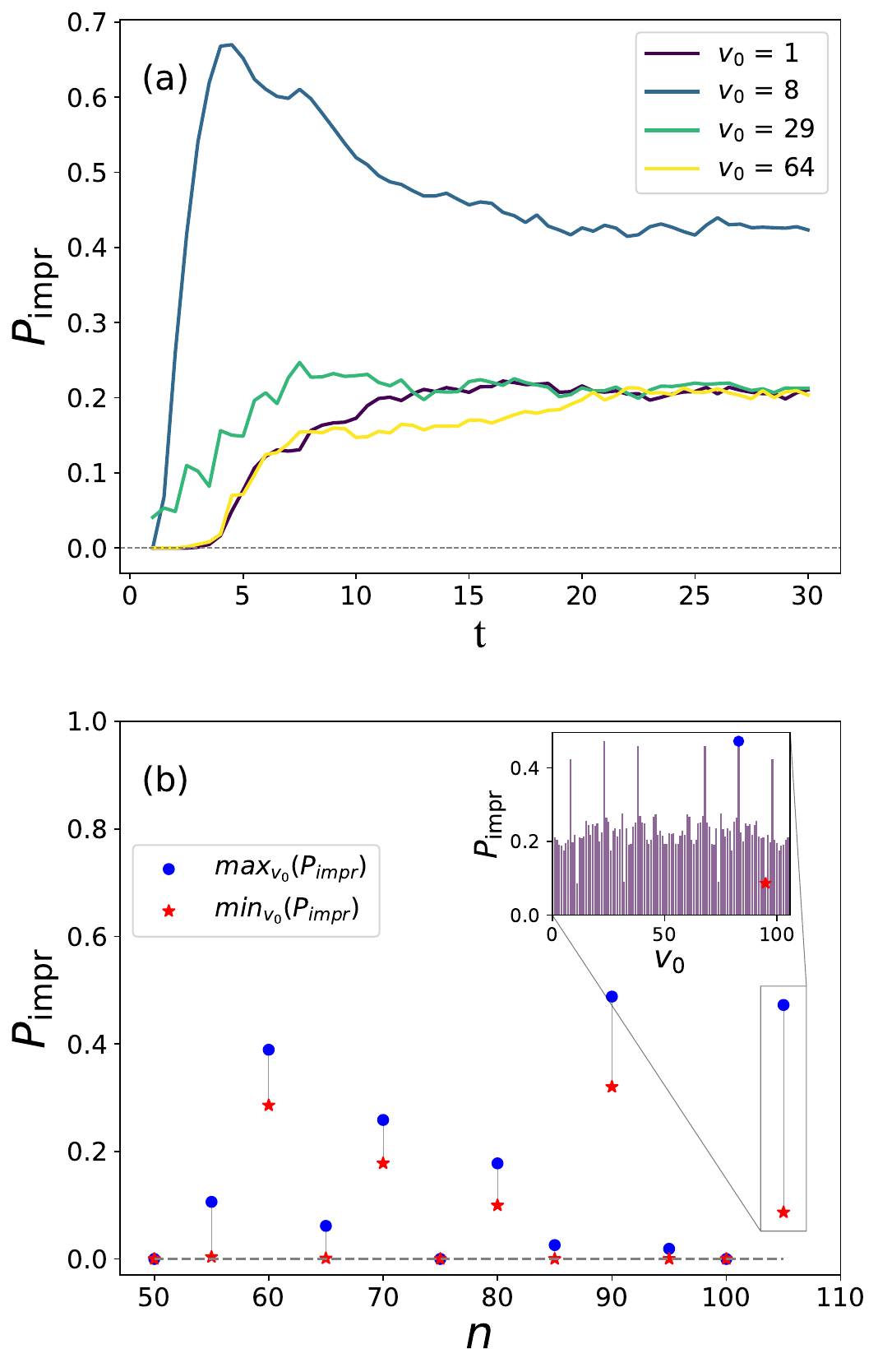}}
\caption{\raggedright 
$P_{\rm{impr}}$ as a function of measurement time $t$  (a), for four different localized initial states, with  $g = 21$ and  $n = 105$; and  as a function of the graph size (b), for $t = 30$ and fixed ratio $n/g = 5$. 
For each value of $n$, blue markers denote the maximum of $P_{\rm impr}$ over all localized initial states $| v_0 \rangle$, while the red stars  represent the corresponding minimum value. The gray line shows the spread between the two values.
For $n = 105$, the inset shows $P_{impr}$ as a function of $v_0$, highlighting the initial states associated with the maximum and minimum values. Probabilities are computed over $10^4$ graph realizations with different  couplings configurations $(j_1, j_2)$ .
\label{fig:coarsed_time_plot}
}

\end{figure}

Finally, Fig.~\ref{fig:coarsed_time_plot} shows that $P_{\rm impr}$ depends on both the measurement time and the initial vertex.
For fixed cluster size, $P_{\rm impr}$ converges to  stationary values at long times, for all choices of $v_0$. 
Hence, the probability that a coarse-grained measurement outperforms a higher-resolution one becomes effectively time-independent in the asymptotic regime. Moreover, the asymptotic value is appreciable for a large range of initial localized  states, couplings configurations and graph sizes (see Fig.~\ref{fig:coarsed_time_plot} (b)).

\section{Conclusions}\label{sec:conclusion}

We have developed an analytical framework for the multiparameter quantum estimation of  coupling strengths in circulant networks probed by continuous-time quantum walks. 
Exploiting the translational symmetry of circulant graphs, we derived closed-form expressions for the quantum Fisher information matrix for arbitrary interaction range and showed that the weak compatibility condition is always satisfied.
As a consequence, the the quantum Cramér-Rao bound is attainable throughout the class of models considered, independently of the number of estimated couplings. 
We further characterize the optimal initial  states that optimize the quantum Cramér--Rao bound and 
analyzed the scaling of the achievable precision with the number of unknown parameters. 
For localized initial states, the quantum Fisher information matrix is diagonal and the corresponding bound grows linearly with the number of estimated parameters. Numerical optimization indicates that this linear scaling is preserved also for optimal probe states, despite the presence of parameter correlations in the corresponding quantum FI matrix.

For the  relevant case of  second-neighbour couplings, we analysed both quantum and classical limits to parameter estimation. We identified the class of graph with  $n=4m$, exhibiting size-independent   QFI matrix and for which both the optimal probe and the resulting estimation bounds acquire a particularly simple structure. In the large-graph limit, all optimal probes converge to a universal form, leading to a universal asymptotic quantum Fisher information matrix and QCR bound.

We then investigated the precision attainable through  position measurements. 
We derived a general analytical expression for the classical Fisher information matrix and showed that its temporal behaviour is determined by the commensurability properties of the populated energy differences. This leads to either periodic or quasi-periodic estimation dynamics. For graphs with $n=4m$ prepared in the optimal probe state, the complete position measurement exhibits an exact periodicity and a Fisher information matrix that is independent of the graph size, implying that the achievable estimation precision is the same for all members of this graph family.

Finally, we studied the effects of finite detector resolution through coarse-grained position measurements. We identified symmetry-induced regimes in which coarse graining completely removes information about the coupling parameters, rendering estimation impossible with QFI-optimal probes. For localized initial states, the coarse-grained FI matrix is periodic in the initial vertex with period equal to the cluster size, thereby reducing the optimization space of the protocol. 
More generally, we found that the dependence of the estimation precision on the measurement resolution is not necessarily monotonic: although coarse graining always degrades performance relative to complete position measurements, lower-resolution measurements can outperform other coarse-grained schemes with finer spatial resolution.

Our results provide a framework and a benchmark for the characterization of larger quantum networks. Natural extensions include graphs with weaker symmetry, where the weak compatibility condition may fail, and the effect of decoherence on the protocols discussed here. More broadly, the interplay between graph symmetry, probe initialization and measurement resolution identified in this work suggests that resolution-aware probe design is an essential ingredient of network tomography with quantum walkers.

\begin{acknowledgments}
The authors thank Marco Genoni and Matteo Paris for useful discussions.
IG and CB are supported by the PRIN 2022 MUR Project EQWALITY (No. 202224BTFZ). CB aknowledges support from Unimi through the Piano di Sostegno alla Ricerca  Linea 2 2025.
\end{acknowledgments}

\appendix

\section{QFI matrix} \label{appendix:qfi}

For an arbitrary initial state in the Hamiltonian eigenbasis, $| \psi_0 \rangle = \sum_{k = 0}^{n - 1} a_k | \epsilon_k \rangle$, with $\sum_{k = 0}^{n - 1} |a_k|^2 = 1$, the state evolved at time $t$ is \begin{align}
    | \psi_t \rangle = \sum_{k = 0}^{n - 1} a_k e^{-it \epsilon_k} | \epsilon_k \rangle.
\end{align}
We assume the system state depends on a vector of $K$ parameters $\mathbf{j} = \left( j_1, j_2, \dots, j_{K} \right)^T$.
Because of the commutativity of the generators $\mathbf{H}_{\alpha}$ of the dynamics, 
\begin{align}
    \ket{\partial _{j_{\alpha}}\psi_t}=-it \mathbf{H}_{\alpha}\ket{\psi_t}.
\end{align}
Substituting into the QFI for pure states, Eq.~\eqref{eq:pure_QFI}, 
one gets:
\begin{align}
   ( \mathbf{F}_Q )_{\alpha, \beta}=4t^2\left[\bra{\psi_t}\mathbf{H}_{\alpha}\mathbf{H}_{\beta}\ket{\psi_t}-\bra{\psi_t}\mathbf{H}_{\alpha}\ket{\psi_t}\bra{\psi_t}\mathbf{H}_{\beta}\ket{\psi_t}\right]
\end{align}
which is equivalent to Eq. \eqref{qficov} since $|\psi_t\rangle=e^{-itH}|\psi_0\rangle$ and $[\mathbf{H},\mathbf{H}_{\alpha}]=0\,\,\forall \alpha$. By exploiting the relations 
\begin{align*}
\mathbf{H}_{\alpha}\,\ket{\epsilon_k}&=-2\cos\left(\frac{2\pi k \,\alpha}{n}\right)\ket{\epsilon_k} \quad \text{ if }\alpha\leq \left\lfloor \frac{n-1}{2}\right\rfloor,\\
   \mathbf{H}_{\frac{n}{2}}\ket{\epsilon_k}&=-(-1)^k\ket{\epsilon_k},
\end{align*}
one gets the elements of the QFI matrix:
\begin{align}
\label{eq:qfi_matrix_incomplete}
&(\mathbf{F}_{Q})_{\alpha, \beta} = 16\, t^2 \left[ \sum_{k = 0}^{n - 1}  |a_k|^2 \cos \left( \frac{2 \pi k}{n} \alpha \right) \cos \left(\frac{2 \pi k}{n} \beta \right) \right. \nonumber \\ 
& \left. - \sum_{k, m = 0}^{n - 1}  |a_k|^2  |a_m|^2  \cos \left(\frac{2 \pi k}{n} \alpha \right)  \cos \left(\frac{2 \pi m}{n} \beta \right)  \right]
\end{align}
if  $\alpha, \beta \leq\left\lfloor \frac{n-1}{2}\right\rfloor$, and 
\begin{align*}
    (\mathbf{F}_{Q})_{\frac{n}{2}, \frac{n}{2}} &= 4\, t^2 \left[ 1 - \left( \sum_{k = 0}^{n - 1} |a_k|^2 (-1)^k \right)^2 \right], \\
    (\mathbf{F}_{Q})_{\alpha, \frac{n}{2}} &= 8\, t^2 \left( \sum_{k = 0}^{n - 1} |a_k|^2 (-1)^k \cos \left(\frac{2 \pi k}{n} \alpha \right) \right.\nonumber \\
    & \left. -\sum_{k, m = 0}^{n - 1}  |a_k|^2  |a_m|^2 (-1)^m \cos \left(\frac{2 \pi k}{n} \alpha \right) \right)
\end{align*}
if  $\alpha \leq\left\lfloor \frac{n-1}{2}\right\rfloor$. The QFI matrix is symmetric $(\mathbf{F}_{Q})_{\alpha, \beta} = (\mathbf{F}_{Q})_{\beta, \alpha}$, for all $\alpha, \beta = 1, 2, \dots, n/2$.

\section{QFI-optimal initial state for two-estimation} \label{appendix:optimal_initial_state}

We compute the optimal initial state for the joint estimation of two coupling strengths $\mathbf{j} =(j_1, j_2)^T$, for a circulant graph with Hamiltonian $\mathbf{H}(2)$, in Eq. \eqref{circH}
As shown in Appendix \ref{appendix:qfi}, for fixed $n$ and time $t$, the QFI matrix, and thus the quantum Cramér-Rao bound $\Tr(\mathbf{F}^{-1}_{Q})$, depends only on the modulus square of the initial state amplitudes. 
Finding an optimal initial state $| \tilde{\psi}_0 \rangle = \sum_{k = 0}^{n - 1} \tilde{a}_k | \epsilon_k \rangle$ means finding the vector $\tilde{\mathbf{a}} = (\tilde{a}_0, \tilde{a}_1, \dots, \tilde{a}_{n - 1})^T$ minimizing  $\Tr(\mathbf{F}^{-1}_{Q})$. 

For the case $K = 2$ at hand, numerical minimization using a sequential least squares programming (SLSQP) algorithm reveals that the optimal probe state depends on the value of $n$ modulo four.
For all values of $n$ considered, the numerical solutions were found to belong to the family of states supported on Hamiltonian eigenstates associated with the extremal eigenspaces of $H_1$ and $H_2$, in agreement with the intuition that  optimal multiparameter probes should draw support from the extremal eigenspaces associated with the individual couplings. Specifically, 
\begin{align}
\label{eq:optimal_initial_state_1}
    | \tilde{\psi}_0 \rangle &= b_1 \left(  | \epsilon_0 \rangle +  | \epsilon_{\frac{n}{2}} \rangle\right) \\
    &+ \left( \sqrt{1 - 2 (b_1)^2 -  (\delta_1)^2} | \epsilon_{ \frac{n}{4} }  \rangle + \delta_1 | \epsilon_{ \frac{3n}{4} }  \rangle \right), \nonumber
\end{align}
if $n = 4k$ for some $k \in \mathbb{N}$,
\begin{align}
    | \tilde{\psi}_0 \rangle & = b_2(n) \left(  | \epsilon_0 \rangle +  | \epsilon_{\frac{n}{2}} \rangle\right) \\
    & + \left(  \sqrt{ \frac{1}{2} - (b_2(n))^2 - (\delta_2)^2 }  | \epsilon_{ \frac{n - 2}{4} } \rangle + \delta_2  | \epsilon_{ \frac{3n + 2}{4} } \rangle  \right) \nonumber \\
		 &+  \left( \sqrt{ \frac{1}{2} - (b_2(n))^2 - (\eta_2)^2 }   | \epsilon_{ \frac{n + 2}{4} } \rangle + \eta_2  | \epsilon_{ \frac{3n - 2}{4} } \rangle  \right), \nonumber
\end{align}
if $n \neq 4k$ for all $k \in \mathbb{N}$, $n$ even, and
\begin{align}
    & | \tilde{\psi}_0 \rangle  = b_3(n) | \epsilon_0 \rangle \\
    &+ \left(  \sqrt{ (c_3(n))^2 - (\delta_3)^2  }  | \epsilon_{ \frac{n - 1}{4} } \rangle + \delta_3  | \epsilon_{ \frac{3n + 1}{4} } \rangle  \right) \nonumber \\
    &+  \left(  \sqrt{  1 - (b_3(n))^2 - (c_3(n))^2 - (\eta_3)^2  }   | \epsilon_{ \frac{n - 1}{2} } \rangle + \eta_3 | \epsilon_{ \frac{n + 1}{2} } \rangle  \right), \nonumber
\end{align}
if $n - 1 = 4k$ for some $k \in \mathbb{N}$,
\begin{align}
\label{eq:optimal_initial_state_4}
    &| \tilde{\psi}_0 \rangle = b_4(n) | \epsilon_0 \rangle \\
    &+ \left(  \sqrt{ (c_4(n))^2 - (\delta_4)^2  } | \epsilon_{ \frac{n + 1}{4} } \rangle + \delta_4  | \epsilon_{ \frac{3n - 1}{4} } \rangle  \right) \nonumber \\
    &+  \left(  \sqrt{  1 - (b_4(n))^2 - (c_4(n))^2 - (\eta_4)^2  }  | \epsilon_{ \frac{n - 1}{2} } \rangle + \eta_4 | \epsilon_{ \frac{n + 1}{2} } \rangle  \right), \nonumber
\end{align}
if $n - 1 \neq 4k$ for all $k \in \mathbb{N}$, $n$ odd.

To gain some intuition as to why the optimal probe states take the forms given in Eqs.~\eqref{eq:optimal_initial_state_1}-\eqref{eq:optimal_initial_state_4}, we note that the  QFI matrix can be rewritten as \begin{equation} \mathrm{Tr}(\mathbf{F}_Q^{-1}) = \frac{C_{11}+C_{22}} {16t^2\left(C_{11}C_{22}-C_{12}^2\right)}, \end{equation} where \[ C_{11} = \mathrm{Var}\!\left[\cos\!\left(\frac{2\pi k}{n}\right)\right], \qquad C_{22} = \mathrm{Var}\!\left[\cos\!\left(\frac{4\pi k}{n}\right)\right], \] and \[ C_{12} = \mathrm{Cov}\!\left[ \cos\!\left(\frac{2\pi k}{n}\right), \cos\!\left(\frac{4\pi k}{n}\right) \right]. \] 
Minimizing $\mathrm{Tr}(\mathbf{F}_Q^{-1})$ therefore  requires simultaneously increasing the variances $C_{11}$ and $C_{22}$ while suppressing the covariance term $C_{12}$.
\newline
For $n=4m$, the indices  $k=0,\, k=\frac{n}{4},\, k=\frac{n}{2},\, k=\frac{3n}{4}$ correspond to the largest spread of the functions entering $C_{11}$ and $C_{22}$.  Moreover, suitable choices of the corresponding amplitudes allow $C_{12}$ to vanish. This observation is consistent with the numerically obtained optimal state in Eq.~\eqref{eq:optimal_initial_state_1}, whose support is restricted to the eigenstates associated with these indices. Similar considerations apply to the remaining remaining classes of $n$ modulo four, leading to the families of optimal states reported above.
\newline
For $n=4$, the coefficient $b_1$ appearing in Eq.~\eqref{eq:optimal_initial_state_1} is independent of the graph size and assumes the numerical value $b_1 \approx 0.588$. In contrast, the coeffcieints $b_i(n)$ for $i = 2,3,4$ depends explicitly on $n$ . Similarly, for odd $n$, the amplitudes $c_i(n)$ are functions of $n$ for $i = 3,4$.
\newline
The parameters $\delta_i$ and $\eta_i$ for $i = 1, \dots, 4$ are arbitrary non-negative numbers (see below), constrained only by the requirement that the square roots in Eqs.~\eqref{eq:optimal_initial_state_1}-\eqref{eq:optimal_initial_state_4} are real.
Consequently,   the quantum Cramér--Rao bound is minimized by an infinite family of probe states, depending not only on the choices of $\delta_i, \eta_i \in \mathbb{R}$, but also on arbitrary phases in the amplitudes. 

In general, the optimal initial probe is a superposition of the non-degenerate Hamiltonian eigenstates, namely$|\epsilon_0 \rangle$ and $|\epsilon_{n/2} \rangle$ for even $n$ (or only $|\epsilon_0 \rangle$ for odd $n$), together with states belonging to the  degenerate eigenspaces spanned by the pairs $| \epsilon_{j} \rangle$ and $| \epsilon_{n - j}\rangle$. 
\newline
The case $n=4$ requires separate treatment. Since the corresponding graph is complete, the optimal state retains the form of Eq.~\eqref{eq:optimal_initial_state_1}, but with $b_1$ replaced by the value $\bar b_1 \approx 0.541$.

\begin{figure}[t]

\centering

{\includegraphics[width=0.9\columnwidth]{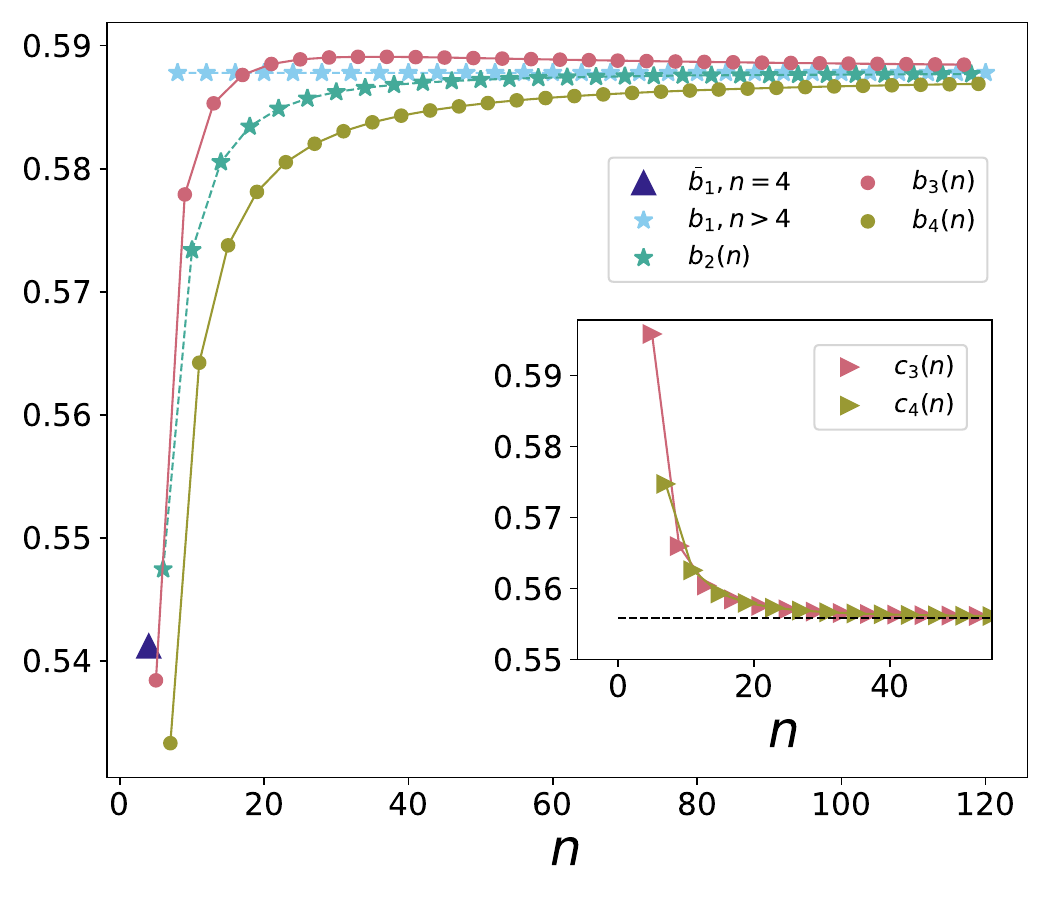}}
\caption{Amplitudes associated with the non-degenerate eigenstates in the optimal probe state $|\tilde{\psi}_0\rangle$, $b_i$, for the simultaneous estimation of two coupling strengths $(j_1,j_2)$, plotted as functions of the number of vertices $n$. 
The blue triangle denotes the value $\bar{b}_1 \approx 0.541$ corresponding to the complete graph. 
For even $n$, the  starred marker represent the coefficients $b_1 \approx 0.588$ and $b_2(n)$ for the case $n = 4k$ and $n \neq 4k$ respectively, $k \in \mathbb{N}$. For odd $n$, the circular-marker  corresponds to the coefficients  $b_3(n)$ and $b_4(n)$ for the cases $n -1 = 4k$ and  $n -1 \neq 4k$ respectively, $k \in \mathbb{N}$. The dashed and solid lines are just a guide for the eye.  All coefficients  $b_i(n)$ converge to the value $b_1$ in the large-$n$ limit. 
Inset: degenerate eigenstate's amplitudes $c_i$ of the optimal initial state, as functions of $n$, for $n - 1 = 4k$ ($c_3(n)$) and $n - 1 \neq 4k$ ($c_4(n)$). In the limit of large $n$, both $c_3(n)$ and $c_4(n)$ converge to the value $\sqrt{1 - 2(b_1)^2} \approx 0.555$ (black dashed line).} \label{fig:initial_amplitudes}
\end{figure}

Fig.~\ref{fig:initial_amplitudes} shows that the coefficients associated to the non-degenerate  eigenstate, $b_i(n)$ for $i = 2,3,4$,  converge to the constant value $b_1$ as the graph size increases. 
Moreover, the inset in Fig.~\ref{fig:initial_amplitudes} shows that the amplitudes associated with the degenerate eigenspaces,  $c_i(n)$ for $i = 3,4$ converge to the constant value $\sqrt{1 - 2(b_1)^2}\approx 0.555$ in the large-$n$ limit.
Therefore, for sufficiently large graphs, the optimal initial states  
do not depend directly on the number of vertices, but only on the residue class of $n$ modulo four.

We use the results in Eqs.~\eqref{eq:optimal_initial_state_1}-\eqref{eq:optimal_initial_state_4} and the general analytical form of the QFI matrix in Eqs.~\eqref{eq:qfi_matrix_incomplete}, to determine the matrix $\mathbf{F}_{Q}$ corresponding to the optimal initial state amplitudes. For even values of $n$, $n = 4k$ for some $k \in \mathbb{N}$, the elements of $\mathbf{F}_{Q}$ are
\begin{align}
\label{eq:QFI_matrix_even_4k}
    (\mathbf{F}_{Q})_{1,1} &=   32\, t^{2} \,b_1^2, \\
    (\mathbf{F}_{Q})_{2,2} &= 128\, t^2\,  (1 - 2\, b_1^2)\,b_1^2, \nonumber\\
    (\mathbf{F}_{Q})_{1,2} &=(\mathbf{F}_{Q})_{2,1} = 0, \nonumber
\end{align}
while for even $n$, $n \neq 4k$ for all $k \in \mathbb{N}$, they are
\begin{align}
\label{eq:QFI_matrix_even_not4k}
    (\mathbf{F}_{Q})_{1,1} &= 16\, t^2 \left(  \left( 1 + \cos\frac{2\pi}{n}  \right)b^2_2(n) + \sin^2 \frac{\pi}{n}  \right)  , \\
    (\mathbf{F}_{Q})_{2,2} &= 128\,t^2 \left(1 - 2\, b^2_2(n)\right) b^2_2(n) \cos^4  \frac{\pi}{n}, \nonumber
 \\
    (\mathbf{F}_{Q})_{1,2} &=(\mathbf{F}_{Q})_{2,1} = 0. \nonumber
\end{align}

For odd $n$, both the $n -1 = 4k$ and $n -1 \neq 4k$ case can be compactly written as
\begin{align}
\makebox[-10pt][l]{\hspace{-10pt}\label{eq:QFI_matrix_odd}}
  (\mathbf{F}_{Q})_{1,1} &= 16\, t^2 \left( - 4\, b_i^4(n) \cos^4\frac{\pi}{2 n} + 2\, b_i^2(n) \cos^2\frac{\pi}{2 n}  \right.\\
  & \times \left(1 + \cos\frac{\pi}{n} - 2\, c_i^2(n) \left(\cos\frac{\pi}{n} \pm \sin\frac{\pi}{2 n}\right) \right)\nonumber\\
  &\left. +\, c_i^2(n)\, (1 - c_i^2(n)) \left(\cos\frac{\pi}{n} \pm \sin\frac{\pi}{2 n} \right)^2 \right), \nonumber
\end{align}
\vspace{3pt}
\begin{align}
\hspace{-100pt} (\mathbf{F}_{Q})_{2,2} &= 16\, t^2 \left(b_i^2(n) \left(1 - \cos^2 \frac{2 \pi}{n} \right) \right. \nonumber \\
   & + c_i^2(n) \cos^2 \frac{\pi}{n} + (1 - c_i^2(n)) \cos^2 \frac{2 \pi}{n}  \nonumber \\
   & - \left[ b_i^2(n) \left(1 - \cos \frac{2 \pi}{n} \right)  \right. \nonumber \\
   & \left. \left.-\, c_i^2(n) \left( \cos \frac{\pi}{n} + \cos \frac{2 \pi}{n} \right) + \cos \frac{2 \pi}{n} \right]^2 \right), \nonumber
\end{align}

\begin{align}
\hspace{-100pt}  (\mathbf{F}_{Q})_{1,2} &= 16\, t^2 \left(2\, c_i^2(n) (1 - c_i^2(n)) \cos^2 \frac{\pi}{2 n} \right. \nonumber\\
    &\times \left(1 - 2 \cos \frac{\pi}{n}\right) \left(\cos \frac{\pi}{n} \pm \sin \frac{\pi}{2 n}\right) \nonumber \\
    &- b_i^4(n) \csc^2 \frac{\pi}{2 n} \sin^4 \frac{\pi}{n} + 2\, b_i^2(n) \cos^2 \frac{\pi}{2 n} \nonumber \\
    &\times \left[ 2 \sin^2 \frac{\pi}{n} + 2\,c_i^2(n) \left(\sin \frac{\pi}{2 n} \mp 1\right) \right. \nonumber \\
    & \left. \left. \times \left(2 \sin \frac{\pi}{2 n} \pm 1 \right) \left(4 \sin^2 \frac{\pi}{2 n} \pm \sin \frac{\pi}{2 n} - 1 \right)  \right] \right), \nonumber \\
   \vspace{20pt}
    (\mathbf{F}_{Q})_{2,1} &= (\mathbf{F}_{Q})_{1,2}, \nonumber
\end{align}
where the upper sign in $\pm$, $\mp$ refers to the case $n - 1 = 4k$, and the lower sign to the case $n - 1 \neq 4k$. The state amplitudes $b_i(n)$, $c_i(n)$ take values $b_3(n)$, $c_3(n)$ and $b_4(n)$, $c_4(n)$ for $n - 1 = 4k$ and $n - 1 \neq 4k$ respectively.

Importantly, none of the matrices Eqs.~\eqref{eq:QFI_matrix_even_4k}-\eqref{eq:QFI_matrix_odd} depend on the arbitrary parameters $\delta_i, \eta_i$, which can thus be set to zero without affecting the value of the QFI matrix. 
Lastly, in the limit of a large number of vertices, $\lim_{n \rightarrow \infty} \sin(x \pi/n) = 0$, $\lim_{n \rightarrow \infty} \cos(x \pi/n) = 1$ $\forall x \in \mathbb{R}$, and the results in Fig.~\ref{fig:initial_amplitudes} show $\lim_{n \rightarrow \infty} b_i(n) = b_1$ for $i = 2,3,4$, and $\lim_{n \rightarrow \infty} c_i(n) = \sqrt{1 - 2(b_1)^2}$ for $i = 3,4$. 
Substituting these limits into Eqs.~\eqref{eq:QFI_matrix_even_4k}--\eqref{eq:QFI_matrix_odd} yields the asymptotic QFI matrix
\begin{equation}
\label{eq:qfi_matrix_limit}
    \mathbf{F}_{Q,\infty} =t^2\, \begin{pmatrix}
32\, b_1^2& 0 \\
0 & 128\,(1 - 2 b_1^2)\, b_1^2 
\end{pmatrix},
\end{equation}
equivalent to Eq.~\eqref{eq:QFI_matrix_even_4k}. 
Thus, in the large-$n$ limit, the optimal QFI matrix and the corresponding   quantum Cramér-Rao bound become independent of the parity of $n$ and of its residue class modulo four.
\newline
For the complete graph case $n = 4$, the elements of the QFI matrix are $(\mathbf{F}_{Q})_{1,1} = 32\, t^{2} \,\bar{b}_1^2$, $(\mathbf{F}_{Q})_{2,2} = 32\, t^2\,  (1 - 2\, \bar{b}_1^2)\,\bar{b}_1^2$, and $(\mathbf{F}_{Q})_{1,2} =(\mathbf{F}_{Q})_{2,1} = 0$.

\section{Classical FI matrix for a complete position measurement} \label{appendix:cfi}

For an arbitrary number $K$ of coupling strengths, we derive the analytical expression for the classical Fisher information matrix Eq.~\eqref{eq:pure_classical_FI}, associated with complete position measurements. We consider the    POVM  $\{ \hat{M}_{s} \}_{s = 1}^n$, with elements $\hat{M}_{s} = | s \rangle \langle s | $ and  $\sum_{s = 1}^{n} \hat{M}_{s} = \hat{\mathbb{I}}_n$. 
We assume that all graph vertices are accessible to the measurement process. 
Assuming the arbitrary initial state in Eq.~\eqref{eq:initial_state},  and using the Fourier representation Eq.~\eqref{eq:eigenstates}, one finds
\begin{align}
\label{eq:scalar_product_position_base_1}
    \langle s | \psi_t  \rangle = \frac{1}{\sqrt{n}} \sum_{k = 0}^{n - 1} a_k e^{-i\left(t\epsilon_k - \frac{2 \pi }{n}(s - 1) k  \right)}
\end{align}
and
\begin{equation}
\label{eq:scalar_product_position_base_2}
  \hspace{-10pt} \langle s | \partial_{\scriptscriptstyle j_{\alpha}} \psi_t \rangle = \frac{2 i t}{\sqrt{n}} \sum_{k = 0}^{n - 1} a_k \cos\!{ \left( \frac{2 \pi k}{n} \alpha \right) } e^{-i\left(t\epsilon_k - \frac{2 \pi(s - 1)k }{n}  \right)}
\end{equation}  
for  $\alpha \le \left\lfloor \frac{n-1}{2}\right\rfloor$,
\begin{equation}
\label{eq:scalar_product_position_base_3}
   \langle s  | \partial_{j_{ \frac{n}{2}}} \psi_t \rangle =  \frac{i t}{\sqrt{n}} \sum_{k = 0}^{n - 1} (-1)^k a_k\,  e^{-i\left(t\,\epsilon_k - \frac{2 \pi (s - 1)k }{n}   \right)},
\end{equation}
for $\alpha=n/2$. It follows that
\begin{align}
\label{eq:FI_matrix_den}
  P_s(\mathbf{j},t&)=\langle  \psi_{t } | \hat{M}_{s} |  \psi_{t }  \rangle = |\langle s | \psi_t \rangle |^2 \\
  &= \frac{1}{n} \sum_{l, k = 0}^{n - 1}  \mathrm{Re}( a^{*}_k a_l) \cos \left(  \theta_{s,l,k}( \mathbf{j}, t)  \right) \nonumber \\
  &\hspace{10 pt} + \frac{1}{n} \sum_{l, k = 0}^{n - 1} \mathrm{Im}( a^{*}_k a_l) \sin \left( \theta_{s,l,k}( \mathbf{j}, t)  \right), \nonumber
\end{align}
where we have defined the angle 
\begin{align}
   \theta_{s,l,k}( \mathbf{j}, t) = t \big(\epsilon_l( \mathbf{j}) - \epsilon_k( \mathbf{j})\big) - \frac{2 \pi}{n} (s - 1) (l - k). 
\end{align}

Equation~\eqref{eq:FI_matrix_den} follows from expanding $|\langle s|\psi_t\rangle|^2$ and expressing the complex exponentials in terms of sine and cosine functions. The contributions $\mathrm{Re}(a_k^{*}a_l)\sin\!\bigl(\theta_{s,l,k}(\mathbf{j},t)\bigr)$ and $\mathrm{Im}(a_k^{*}a_l)\cos\!\bigl(\theta_{s,l,k}(\mathbf{j},t)\bigr)$ are antisymmetric under the exchange $k\leftrightarrow l$ and therefore vanish upon summation over $k$ and $l$. Consequently, $|\langle s|\psi_t\rangle|^2$ is real for all initial-state amplitude vectors $\mathbf{a}$, all vertices $s=1,\ldots,n$, and all  times $t$.
From Eqs~\eqref{eq:scalar_product_position_base_1}-\eqref{eq:scalar_product_position_base_2},
\begin{align}
\label{eq:FI_matrix_num}
    N_{\alpha,s}(\mathbf{j},t)&=\mathrm{Re}  \left( \langle \psi_{t } | \hat{M}_{s} | \partial_{j_{\alpha}}\psi_{t}\rangle \right) \\
    &=  \mathrm{Re} \left( \langle \psi_t | s \rangle \langle s | \partial_{j_{\alpha}} \psi_t \rangle \right) \nonumber\\
    &= \frac{2t}{n} \sum_{l, k = 0}^{n - 1} \cos \left( \frac{2\pi l}{n} \alpha \right)   \mathrm{Re}( a^{*}_k a_l) \sin \left(  \theta_{s,l,k}( \mathbf{j}, t)  \right) \nonumber\\
    &- \frac{2t}{n} \sum_{l, k = 0}^{n - 1} \cos \left( \frac{2\pi l}{n} \alpha \right)   \mathrm{Im}( a^{*}_k a_l) \cos \left( \theta_{s,l,k}( \mathbf{j}, t)  \right) \nonumber
\end{align}
for  $\alpha \le \left\lfloor \frac{n-1}{2}\right\rfloor$, and
\begin{align}
   N_{n/2,s}(\mathbf{j},t)& =\mathrm{Re}  \left( \langle \psi_{t } | \hat{M}_{s} | \partial_{j_{\frac{n}{2}}}\psi_{t}\rangle \right) \\
    &= \frac{t}{n} \sum_{l, k = 0}^{n - 1}(-1)^l \mathrm{Re}( a^{*}_k a_l) \sin \left(  \theta_{s,l,k}( \mathbf{j}, t)  \right) \nonumber\\
    &- \frac{t}{n} \sum_{l, k = 0}^{n - 1} (-1)^l  \mathrm{Im}( a^{*}_k a_l) \cos \left( \theta_{s,l,k}( \mathbf{j}, t)  \right) \nonumber
\end{align}
for $\alpha = n/2$. 
Substituting these equations into Eq.~\eqref{eq:pure_classical_FI} yields the analytical expression for the classical Fisher information matrix associated with complete position measurements, Eq. \eqref{eq:classicalFIM}.
\newline
\newline
As an example, we report the explicit expression of Eq. \eqref{eq:FI_matrix_den} and \eqref{eq:FI_matrix_num}, for the estimation of two coupling strengths $\mathbf{j} = (j_1, j_2)$,  for an optimal initial state with number of vertices $n = 4k$, $k \in \mathbb{N}$. From state $| \tilde{\psi}_0 \rangle$ in Eq.~\eqref{eq:optimal_initial_state_1},
\begin{align}
\label{eq:probability_n4k}
 P_{s}(\mathbf{j},t)   &=  \langle  \psi_{t } | \hat{M}_{s} |  \psi_{t }  \rangle \\
 &= \frac{1}{n} \left( 1 - 2\, b_1^2 \cos \left( \pi s - 4 j_1 t \right) \right.\nonumber \\
      &  + 2\, b_1 \sqrt{1 - 2\, b_1^2} \sin \left( \frac{\pi s}{2} - 2 (j_1 - 2 j_2) t \right)  \nonumber \\
      & \left. + 2\, b_1 \sqrt{1 - 2\, b_1^2}\sin \left( \frac{\pi s}{2} - 2 (j_1 + 2 j_2) t \right) \right), \nonumber
\end{align}
and
\begin{align}
\label{eq:j1_numerator}
      N_{1,s}(\mathbf{j},t)&=  \mathrm{Re} 
      \left( \langle \psi_{t } | \hat{M}_{s} | \partial_{j_{1}}\psi_{t}\rangle \right) =\\
      &-\frac{4}{n} \, t \, \cos \left( \frac{\pi s}{2} - 2 j_1 t \right) \times\nonumber\\
        &\left( b_1 \sqrt{1 - 2\, b_1^2} \cos(4\, j_2\, t) + 2\, b_1^2 \sin \left( \frac{\pi s}{2} - 2 j_1 t \right) \right),\nonumber
\end{align}
\begin{align}
\label{eq:j2_numerator}
     N_{2,s}(\mathbf{j},t)&=   \mathrm{Re}  \left( \langle \psi_{t } | \hat{M}_{s} | \partial_{j_{2}}\psi_{t}\rangle \right) = \\
        &-\frac{8}{n}\,t \, b_1 \sqrt{1 - 2\, b_1^2}  \sin(4 j_2 t) \, \sin \left( \frac{\pi s}{2} - 2 j_1 t \right). \nonumber
\end{align}
\begin{figure}[t]
\centering
{\includegraphics[width=0.9\columnwidth]{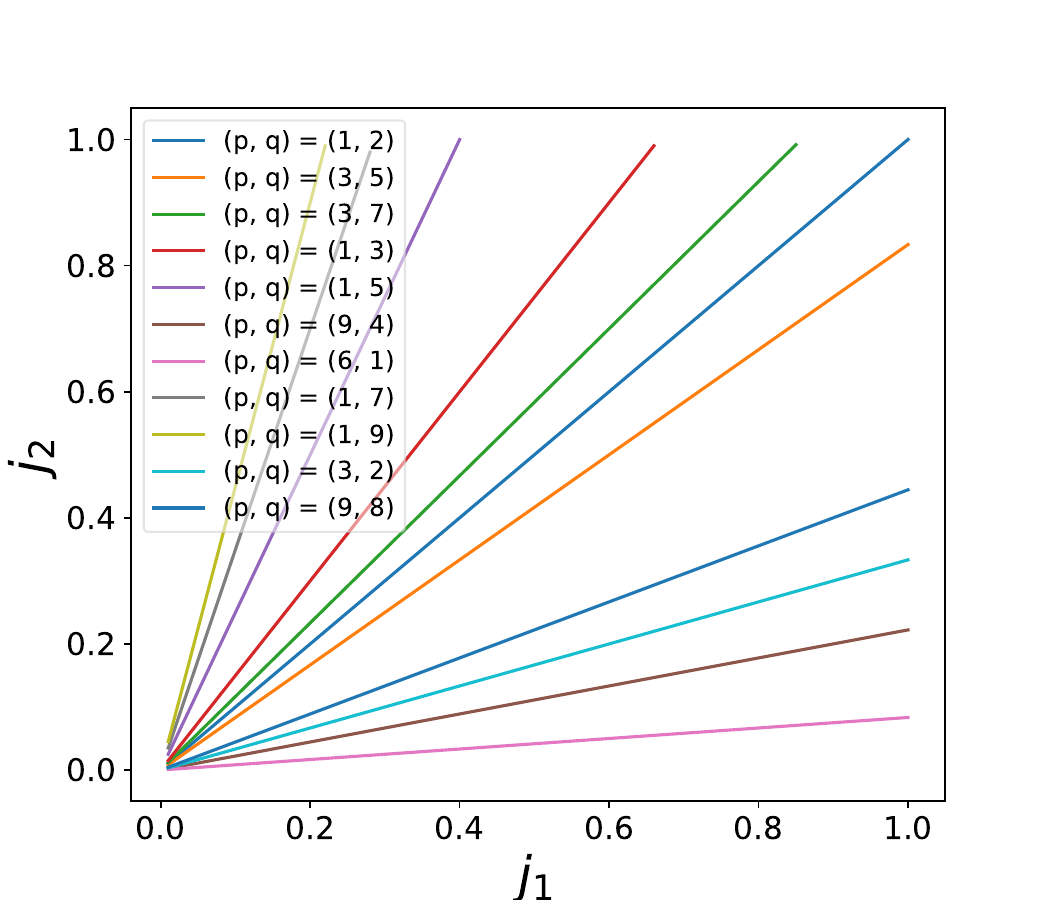}}
\caption{The plotted lines correspond to coupling strength ratios $j_1/(2j_2) = p/q$, with coprime integers $(p, q)$, for which the rescaled classical Cram\'er--Rao bound is exactly periodic in time with period $T = \pi p /j_1 = \pi q /(2 j_2)$, under the condition $n = 4m$, $m \in \mathbb{N}$. Ten such pairs $(p,q)$, $p, q < 10$, are chosen at random and shown in the figure.  } \label{fig:commensurate_couples}
\end{figure}
Inspection of Eqs.~\eqref{eq:probability_n4k}--\eqref{eq:j2_numerator} shows that all trigonometric functions entering the classical FI matrix depend on the two fundamental frequencies: 
\begin{equation}
    \omega_1 = 2 j_1, \qquad \omega_2 = 4 j_2.
\end{equation}
Indeed, all  frequencies appearing in the sine and cosine functions can be written as linear combinations of $\omega_1$ and $\omega_2$.
Consequently, the rescaled matrix $\mathbf{F}_{C}/t^2$ is periodic in time if and only if the frequencies $\omega_1$ and $\omega_2$ are commensurate, i.e. 
$     \frac{\omega_1}{\omega_2} = \frac{p}{q} \in\mathbb Q$ where $p$ and $q$ are coprime integers,
which implies:
\begin{equation} \label{eq:commensurate_condition} \frac{j_1}{2j_2} = \frac{p}{q} \in\mathbb Q, \end{equation}
see Figure \ref{fig:commensurate_couples}.
In this case, there exists a common  period $T$  satisfying
\begin{align}
\label{eq:periods}
    \omega_1 T = 2 \pi p, \qquad \omega_2 T = 2 \pi q.
\end{align}
The same periodicity property is inherited by the rescaled classical Cramér--Rao bound $t^2\Tr(\mathbf{F}_C^{-1})$ since, for a $2\times2$ matrix,  
\begin{align}
    \Tr(\mathbf{F}_C^{-1}) = \frac{\Tr(\mathbf{F}_C)} {\det(\mathbf{F}_C)}.
\end{align} 
If $j_1, j_2$ are rational numbers, the commensurability condition in Eq.\eqref{eq:commensurate_condition} is automatically satisfied. 
For any  pair  $(j_1, j_2)$ obeying Eq.\eqref{eq:commensurate_condition} with $p$ and $q$ coprime integers, the fundamental period, from Eq.~\eqref{eq:periods}, is  $T = \pi p /j_1 = \pi q /(2 j_2)$. 
Furthermore, if the couplings can be written as $j_1 = p d $ and $2 j_2 = q d $, where $d=\gcd(j_1,2j_2)$ is identified with the greatest common divisor of the commensurable real numbers $j_1$ and $2j_2$, then  the  period can be written as
\begin{align}
    T = \frac{\pi}{d}=\frac{\pi}{\gcd(j_1,2 j_2)}.
\end{align}
As an example,  for $j_1 = 1$, $j_2 = 1/(2\sqrt{2})$, one obtains $j_1/(2 j_2) = \sqrt{2}$ which is irrational. In this case no common period exists, and the rescaled classical Cramér--Rao bound is quasi-periodic rather than exactly periodic.

\section{Coarse-grained FI matrix for a QFI-optimal probe state with $n = 4m$} \label{appendix:coarse_grained}
In this appendix we show that the coarse-grained Fisher information matrix vanishes identically when the graph size is $n=4m$, the cluster size is $g=4m'$, with $m,m'\in\mathbb{N}$, and the probe is prepared in a QFI-optimal state.

The matrix elements of $\mathbf{F}^{g}_{C}$ are given by Eq.~\eqref{eq:coarse_grained_FI}, whose numerators  contain the derivatives 
$\partial_{j_\alpha}P \left( C_g | ~ \mathbf{j} \right)=\sum_{x\in C_g}\partial_{j_\alpha} P_x(\mathbf{j},t)$ 
(see Eq.~\eqref{eq:numerator_FI_coarsed}).
For QFI-optimal probe states, the single-site probabilities satisfy the period-four relation $ P_{x + 4}(\mathbf{j},t)= P_x(\mathbf{j},t)$ for all times. Therefore, for  the ordered clusters
$C_k=\{k,\dots,k+g-1\}$ with $g=4m'$ the derivatives satisfy
\begin{equation}
    \partial_{j_\alpha}P \left( C_k | ~ \mathbf{j} \right)=m'\,\sum_{x=k}^{k+3}\partial_{j_\alpha}P_x(\mathbf{j},t).
\end{equation}
Using Eq.~\eqref{eq:j1_numerator}, one finds
\begin{equation}
\partial_{j_1}P_x(\mathbf{j},t)=f_1(n,t,j_2)\cos\theta_x+\tfrac{1}{2}f_2(n,t)\sin 2\theta_x,
\end{equation}
where $\theta_x=\pi x/2-2j_1t$, $f_1=-\tfrac{8}{n}\,t\,b_1\sqrt{1-2b_1^2}\cos(4j_2t)$ and $f_2=-\tfrac{16}{n}\,t\,b_1^2$. 
Because $\theta_{x+2}=\theta_x+\pi$, the cosine contributions cancel pairwise over any four consecutive sites. Likewise, $2\theta_{x+1}=2\theta_x+\pi$, implies pairwise cancellation of the sine contributions. 
Hence $\sum_{x=k}^{k+3}\partial_{j_1} P_x(\mathbf{j},t)=0$ for every $k$, and so $\partial_{j_1}P \left( C_k | ~ \mathbf{j} \right)=0$. The same argument applied to Eq.~\eqref{eq:j2_numerator} gives $\partial_{j_2}P \left( C_k | ~ \mathbf{j} \right)=0$. Therefore all numerators in Eq.~\eqref{eq:coarse_grained_FI} vanish, implying $(\mathbf{F}^{g}_{C})_{\alpha,\beta}=0$ for $\alpha,\beta=1,2$.
The coarse-grained probabilities are independent of the couplings.
Since the clusters form a partition of the graph and all have the same size, it follows from normalization that 
\begin{equation} 
P \left( C_k \middle| \mathbf{j} \right)=\frac{g}{n} 
\end{equation} 
for every cluster $C_k$.

\bibliography{bibliography}

\end{document}